# Enhancing one-dimensional Particle-in-Cell simulations to self-consistently resolve instability-induced electron transport in Hall thrusters

F. Faraji [1†], M. Reza [1], A. Knoll [1]

[1]Plasma Propulsion Laboratory, Department of Aeronautics, Imperial College London, Exhibition Road, London, SW7 2AZ, United Kingdom

**Abstract**: The advent of high-power Hall thrusters and the increasing interest towards their use as a primary propulsion system for various missions have given a new boost to the efforts aiming at self-consistent predictive modeling of this thruster technology. In this article, we present a novel approach, which allows enhancing the predictive capability of one-dimensional Particle-in-Cell (PIC) simulations to self-consistently capture the wave-induced electron transport due to the azimuthal instabilities in Hall thrusters. The so-called "pseudo-2D" PIC scheme resulting from this approach is extensively tested in several operating conditions. The results are compared against a well-established 2D3V axial-azimuthal reference case in terms of the axial profiles of the time-averaged plasma properties, the azimuthal electric field fluctuations and their dispersion features, and the contributions of the force terms in the electron azimuthal momentum equation to the cross-field mobility. We have demonstrated that the pseudo-2D PIC provides a prediction of the above aspects that compares very closely in almost all conditions with those from the full-2D simulation. In addition, the sensitivity of the pseudo-2D simulation results to the numerical parameters associated with our approach is assessed in detail. The outcomes of these analyses have casted light on the next steps to further improve the approach.

## Section 1: Introduction

Hall thruster technology today is the propulsion option of interest for a broad spectrum of space mission scenarios. Hall thrusters are highly efficient devices that provide many advantages in terms of power-to-thrust ratio, lifetime, manufacturing simplicity, and operational versatility compared to other electrostatic EP devices. With the emergence of novel near-Earth and deep-space mission scenarios, the interest in the development of Hall thrusters with power levels above 5 kW has surged in the recent years. Nevertheless, as the power level and the total throughput capability has increased, the necessity of a self-consistent numerical design tool has been more felt for reliable and accurate prediction of thrusters' performance and for carrying out parametric studies, typical to the preliminary design phases of the thruster, at a reasonably low computational cost. Indeed, being the current development approach of Hall thrusters heavily test-dependent, the incurred cost and required time to qualify, for instance a very high-power 20 kW Hall thruster, is enormous. In addition, the life testing of Hall thrusters, which amounts to continuous firing times on the order of tens of thousands of hours, is truly challenging from the financial and scheduling aspects [1]. Finally, the background pressure and electrical effects of vacuum chambers on performance and behavior of Hall thrusters introduce uncertainties in the results of on-ground performance characterizations and the extent of similarity with the actual thruster performance on-orbit [1]. All these issues become much more serious for high-power Hall thrusters [2].

A major hurdle to achieve a design-aiding tool roots in the physics of Hall thrusters' operation that is not yet fully understood. One of the main open questions regarding the underlying physics of these devices pertains to understanding the electrons' dynamics, and particularly, the mechanisms that render the rate of cross-magnetic-field diffusion much larger than those predicted by the classical collision theory [3]. This phenomenon, referred to as "anomalous" electron cross-field transport, poses a real challenge to the development of self-consistent simulation models. This is because, in principle, the mechanisms behind this phenomenon and the significance of their contribution to cross-field transport can vary temporally within a period of thruster's characteristic current oscillation, spatially from the near-anode to near-plume region, and from one operating condition to another [3][4].

In view of these complexities, high-fidelity, though computationally costly, kinetic Particle-in-Cell (PIC) simulations are being extensively employed to investigate the electrons' dynamics and interactions in order to provide foundational physical insights that can in turn enable describing the electron transport problem through closure models [5]. It is worth noting that the computational cost of traditional PIC codes is such that their direct application as a predictive design and test-aiding tool for full-scale Hall thrusters is currently impractical. Hence, efforts have been dedicated in parallel to advance fully-fluid and hybrid fluid-PIC simulations that would implement the derived models from kinetic simulations to close the system of equations for the electron flow

---

[†]Corresponding Author (f.faraji20@imperial.ac.uk)

[6][7][8]. In theory, if one succeeds in establishing a rigorous and generalizable closure model, the relatively lower computational cost of hybrid and fully-fluid simulations warrant their application as an engineering numerical tool. Nevertheless, describing electrons as a fluid in Hall thrusters cannot be completely justified and is at best a simplifying approximation. This is because of two reasons: first, due to the E×B configuration of Hall thrusters, there is a high degree of anisotropy leading to a non-Maxwellian distribution function. The deviation from a Maxwellian distribution is especially true in the acceleration region of the thrusters [9], which is the most crucial with respect to performance. As the fluid description does not resolve the evolution of the distribution function, the effects on the global plasma behavior, and hence performance, due to variations in electrons' distribution cannot be readily captured. Second, it has been shown recently [10][11] that the nonlinear convective inertia and viscous terms, usually neglected in fluid descriptions of electrons, can play an important role in transport. Therefore, the effects of these missing physics need to be reflected in the closure models for fluid/hybrid codes. This necessity can introduce an incompatibility with the transport models derived based on observations in PIC/kinetic simulations. Finally, the multiscale, multidimensional nature of plasma phenomena in Hall thrusters can be comprehensively captured only using 3D PIC simulations, which are currently unattainable [12].

Noting these limitations, we propose a novel approach to enable the computationally efficient and high-fidelity one-dimensional PIC simulations capturing self consistently the contribution of azimuthal instabilities to electron transport, a capability that has been exclusive so far to multi-dimensional PIC simulations resolving the azimuthal coordinate of the problem [12]. For a 2D domain, this is done by introducing an approximation of the potential field, which leads to a "pseudo-2D" description of the problem. In the present article, we describe an implementation of the pseudo-2D scheme, which we term "double region", that is supported by recent observations in Hall thrusters concerning the axial evolution of the main characteristics of azimuthal instabilities. As a special case of this implementation, we also discuss the so-called "single-region" formulation. We present the results from the single- and double-region pseudo-2D simulations in a set of operating conditions, comparing the observations with the results of a full-2D axial-azimuthal benchmark case [13]. We also discuss the sensitivity analyses we have performed on the numerical parameters involved in the pseudo-2D scheme.

The rest of the paper is organized as follows: in Section 2, the main features of the Particle-in-Cell code that has been developed by the authors in Imperial Plasma Propulsion Laboratory (IPPL) is introduced. In this section, we also present the adopted problem formulation that leads to the pseudo-2D PIC scheme. Section 3 provides an overview of the axial-azimuthal simulation setup used in our study and compares the pseudo-2D simulation results with those from a full-2D simulation in a reference benchmark case. The comparison between the predictions of the pseudo-2D and full-2D simulations are extended to a range of operating conditions in Section 4. Section 5 is dedicated to the discussion of the sensitivity analyses performed on three parameters associated with the pseudo-2D PIC scheme: the weighting coefficients used to approximate the 2D potential distribution as the superimposition of two 1D potential fields, the azimuthal length of the simulation domain, and the location of the boundary between the two regions in the double-region implementation. Finally, we highlight the main findings and outcomes of this research as the conclusion in Section 6 and outline the next steps aimed at improving the pseudo-2D PIC scheme and its underlying formulation.

## Section 2: IPPL Particle-in-Cell code and the pseudo-2D PIC scheme

We have developed an electrostatic explicit Particle-in-Cell code with the specific aim of being used to study the plasma behavior in a high-power magnetically shielded Hall thruster. As such, the code is equipped with several modules that resolve the physical aspects of importance for the operation of these devices. Particular attention is, thus, dedicated to the functions capturing inter-particle collisions and plasma-wall interactions. The collision module is based on the Monte Carlo (MC) scheme [14] for the electrons and ions interaction with neutrals. A Direct Simulation Monte Carlo (DSMC) subroutine [15] is included to capture neutral-neutral collisions, ionization of singly charged ions, and interaction of multiply charged ions with neutral particles. The wall interaction module incorporates multiple schemes for Secondary Electron Emission (SEE) from the thruster's channel walls, including a simple linear model [16] and a more advanced Monte-Carlo-based model using Vaughan formulation [17].

The IPPL code is written in Julia language [18] and uses the built-in Julia Random Number Generator (RNG) that follows the Xoshiro256++ algorithm [19] by default. Several algorithms to solve electric potential have been implemented, from the Thomas tridiagonal to the Pre-conditioned Conjugate Gradient (PCG) [20]. A linear weighting algorithm (also referred to as the Cloud-in-Cell approach [21]) is pursued to scatter particle-based information, such as the charge, to the grid points and to gather grid-based data, like the electric field, onto the particles' location. The particle push algorithm is based on the classic leap-frog scheme [22], with the push function for the magnetized electrons following the Boris method [23].

Since our PIC code is recently developed, its verification against rigorous benchmark cases at both the module and global level has been crucially important. The code's extensive verification campaign is successfully concluded, confirming the correct implementation the core modules and the additional-physics subroutines (such as collisions and wall interactions). We present the results from two of these benchmarking activities in the Appendix. In this regard, Section A.1 presents the results of the code's verification against the 1D Capacitively Coupled Discharge benchmark [24], and Section A.2 shows the verification results of the code in a 1D azimuthal setup similar to that adopted in Ref. [25].

## 2.1. Description of the Pseudo-2D PIC scheme

Referring to Figure 1, we start by separately decomposing a 2D computational domain into one horizontal region ($\Omega^x$) along the x (axial) direction and two vertical regions ($\Omega_m^y$) along the y (azimuthal) direction, with $m \in [1,2]$. As there are two vertical regions, we call this implementation the "double region". At the intersection of the axial region with each of the vertical ones, we assume that the 2D potential field ($\phi(x,y)$) can be expressed in terms of a linear combination of two independent potential fields $\phi^x$ and $\phi_m^y$, where $\phi^x$ is the potential field in region $\Omega^x$, and $\phi_m^y$ is the potential field in each region $\Omega_m^y$. Thus, we write

$$\phi(x,y) = C^x \phi^x + C_m^y \phi_m^y, \qquad \text{with} \quad x \in \Omega^x \ , \ y \in \Omega_m^y \ \& \ m \in [1,2]. \tag{Eq. 1}$$

In the above equations, $C^x$ and $C_m^y$ are the weighting coefficients used to approximate the 2D potential field, with $C^x$ being constant in $\Omega^x$, and each coefficient $C_m^y$ being constant in its corresponding $\Omega_m^y$ region.

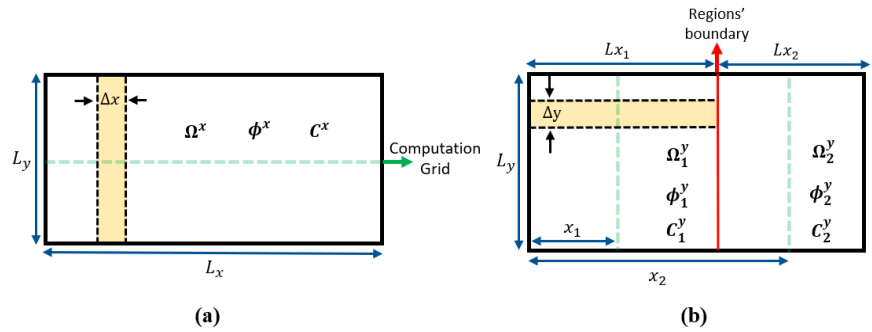

Figure 1: Schematics of the computational domain and the defined "Regions"; (a) the horizontal region ($\Omega^x$), and (b) the vertical regions ($\Omega_1^y$ & $\Omega_2^y$)

At this point, we point out the recent evidence from full-2D axial-azimuthal simulations [28] that, along the axial direction of a Hall thruster, there is a transition at the location of the ion sonic point where the characteristics of the azimuthal waves change from short wavelength to long wavelength oscillations, with an associated drop in frequency. However, before and after the transition point, the wave's main features (frequency, wave number and amplitude) remain relatively constant along the axis. Although no rigorous physical explanation yet exists for this observation, it supports defining the azimuthal potential function $\phi_m^y$ in each region $\Omega_m^y$ as being only a function of the y-coordinate. Similarly, we assume that, within the region $\Omega^x$, $\phi^x$ is, by definition, only a function of the x-coordinate. Therefore, the 1D approximation is assumed for $\phi^x$ and $\phi_m^y$.

In this regard, it is noted that, in a general case, the vertical and horizontal regions are to be defined according to the involved physics of the problem such that, within each region, the 1D approximation would be valid for the corresponding potential fields ($\phi^x$ and $\phi^y$).

Based on the definition of $\phi^x$ and $\phi_m^y$ as above, the Poisson's equations for these potential fields can be written as

$$\frac{d^2\phi^x(x)}{dx^2} = -\frac{1}{\epsilon_0}\frac{1}{L_y}\int_0^{L_y}\rho(x,y)\ dy = -\frac{\rho^x(x)}{\epsilon_0}, \tag{Eq. 2}$$

and,

$$\frac{d^2\phi_m^y(y)}{dy^2} = -\frac{1}{\epsilon_0}\frac{1}{L_{x_m}}\int_{x_m-\frac{L_{x_m}}{2}}^{x_m+\frac{L_{x_m}}{2}} \rho(x,y)\ dx = -\frac{\rho_m^y(y)}{\epsilon_0},$$ (Eq. 3)

for computational cells extending over the entire width of the regions (highlighted areas in Figure 1(a) and (b)). In Eqs. 2 and 3, $L_y$ is the extent of the domain along the azimuthal coordinate, $L_{x_m}$ is the axial extent of each region $\Omega_m^y$, $\epsilon_0$ is the permittivity of free space, and $x_m$ is the location of the computation grid inside each region $\Omega_m^y$ with respect to the origin. $\rho(x,y)$ is the 2D charge density distribution, and $\rho^x$ and $\rho_m^y$ are, respectively, its average over the width of the regions $\Omega^x$ and $\Omega_m^y$.

At this point, we perform a change of variables in Eq. 1 and define

$$\zeta(x) = C^x\phi^x(x),$$ (Eq. 4)

$$\eta_m(y) = C_m^y\phi_m^y(y),$$ (Eq. 5)

such that, $\phi(x,y) = \zeta(x) + \eta_m(y)$ at the regions' intersection. Accordingly, rewriting Eqs. 2 and 3 using these definitions, we have

$$\frac{d^2\zeta(x)}{dx^2} = -C^x\frac{\rho^x(x)}{\epsilon_0},$$ (Eq. 6)

$$\frac{d^2\eta_m(y)}{dy^2} = -C_m^y\frac{\rho_m^y(y)}{\epsilon_0}.$$ (Eq. 7)

Concerning the weighting coefficients $C^x$, $C_1^y$ and $C_2^y$, we assumed their values to be the same and equal to 0.5 for the preliminary proof-of-concept studies presented in this paper. We discuss the sensitivity of the simulation results to the value of these coefficients in Section 5. Eqs. 6 and 7 can be solved as a decoupled system of 1D ODEs for $\zeta(x)$ and $\eta_m(y)$. In turn, the electric field is obtained using $\vec{E} = -\vec{\nabla}\phi(x,y) = -\vec{\nabla}(\zeta(x) + \eta_m(y))$.

Implementing the above approach to solve the potential and the electric field leads to a pseudo-2D PIC scheme since it effectively amounts to solving in-parallel 1D PIC simulations along the axial and azimuthal coordinates of the thruster with the simulations sharing the same macroparticles. Therefore, the total number of required macroparticles is essentially on the same order of a 1D simulation, and so too is the computational cost. We provide a quantitative comparison between the necessary computational resources for a pseudo-2D versus a full-2D PIC simulation in Section 3.2.

A special case of the above formulation is obtained by reducing the number of vertical regions from two to one. We refer to this case as the "single-region" implementation. The single-region case implies that the characteristics of the excited waves along the azimuthal coordinate correspond to the axially averaged plasma properties. Even though this is not the best approximation in case of a Hall thruster, we performed the single-region pseudo-2D simulations as well to assess its capabilities in capturing the essential azimuthal physics and in incorporating the average effect of the instabilities in electrons' cross-field mobility.

It is important to underline that we focus in this work on the single- and double-region implementations of the pseudo-2D PIC scheme since the main aim was to preliminarily verify the approach. Nonetheless, the pseudo-2D scheme does not have an inherent limitation on the number of regions and, indeed, increasing the number of regions $\Omega_m^y$ along the axial direction can allow resolving possible axial variations that may exist in the characteristics of the azimuthal waves, as it has been reported in some other self-consistent PIC simulations [11]. However, we decided to leave such studies to the future work when the underlying formulation is mathematically more mature.

## Section 3: Description of the simulation setup and results for the nominal benchmark

Pseudo-2D axial-azimuthal (x-y) simulations were performed for the conditions and setup of a full-2D x-y benchmark [13]. The operating condition used in this benchmark publication is referred to here as the "nominal". In this section, we focus on comparing the pseudo-2D versus the full-2D simulation results for the nominal case in terms of the time-averaged plasma properties. A more detailed comparison between the predictions of the two simulation schemes is presented in Section 4. In that section, we expand the analyses to include the dispersion features of the observed oscillations and the contribution of the azimuthal momentum terms to electron transport for a set of operating conditions comprising the nominal one.

### 3.1. Simulation domain and conditions

The simulation domain length in the axial direction is 2.5 cm, covering 0.75 cm of a Hall thruster's channel and 1.25 cm of the plume region. Therefore, the near-anode region is not included. Initially, electron and ion particles are loaded with a density of $5 \times 10^{16}$ m$^{-3}$, uniformly throughout the simulation domain at exactly same locations. Both particle species are sampled from a Maxwellian distribution function at 10 eV for electrons and 0.5 eV for ions.

The axial profile of the radial magnetic field follows the same Gaussian profile as that adopted in [13], and it is illustrated in Figure 2(a). The magnetic field peak is at the exit plane which corresponds to x = 0.75 cm. Similar to Ref. [13], the plasma is assumed to be collisionless and, thus, the ionization process is assimilated (as in [13]) using an imposed temporally invariant source with a cosine-shaped profile along the x-direction (Figure 2(b)). The peak of the ionization source is determined such that the maximum ion current density will be 400 A/m$^2$.

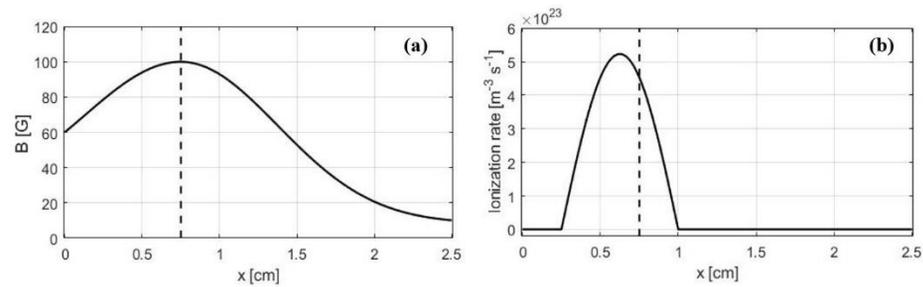

Figure 2: Axial profiles of (a) the radial magnetic field, and (b) the imposed ionization source; the dashed black line represents the location of the exit plane and the peak intensity of the magnetic field.

The Dirichlet potential boundary conditions for the Poisson solver in the axial direction are 200 V at x = 0 (domain's anode side) and 0 V at x = 2.5 (domain's cathode side). In the azimuthal direction, a periodic boundary condition with zero-volt reference potential has been applied to both ends. The Thomas tridiagonal algorithm is used to solve the potential in the axial and azimuthal directions.

With regards to the particle boundary conditions, both ions and electrons reaching either the anode or the cathode side are removed from the simulation. However, to maintain the discharge, it is necessary to constantly inject electrons into the domain from the cathode side. The number of electrons to inject is obtained using the current-continuity approach described in [13]. These electrons are sampled from a Maxwellian distribution at 10 eV. The azimuthal direction is periodic, and the particles leaving the domain at one end are injected back from the other end.

Table 1 summarizes the values of the main parameters used in our simulations throughout this paper. It is noteworthy that the chosen number of initial macroparticles per cell, which can affect the results especially in the azimuthal direction, are in line with those used in the benchmark case [13]. In addition, the baseline azimuthal length of the domain is 0.5 cm. However, as the wavelength of the azimuthal instabilities of interest is proportionate to the number density through the Debye length [3], for the low-current-density simulation cases of Section 4, the azimuthal extent was increased to 1 cm to ensure that the simulation can capture a few instability wavelengths.

| Parameter | Value [unit] |
|---|---|
| Computational Parameters | |
| Time step ($\Delta t$) | $5 \times 10^{-12}$ [s] |
| Total simulation time ($t_{final}$) | $20 \times 10^{-6}$ [s] |
| Axial domain length ($L_x$) | 2.5 [cm] |
| Azimuthal domain length ($L_y$) | 0.5/1.0 [cm] |
| Cell size ($\Delta x = \Delta y$) | $5 \times 10^{-3}$ [cm] |
| Initial number of macroparticles per cells for axial grid ($N_{ppc_x}$) | 150 |
| Initial number of macroparticles per cells for azimuthal grid ($N_{ppc_y}$) | 375 |
| Physical Parameters | |
| Initial plasma density ($n_{p,init}$) | $5 \times 10^{16}$ [m$^{-3}$] |
| Electron injection temperature ($T_e$) | 10 [eV] |
| Ion injection temperature ($T_i$) | 0.5 [eV] |
| Discharge Voltage ($V_d$) | 200 [V] |

Table 1: Summary of the numerical and physical parameters used for the pseudo-2D simulations

### 3.2. Predictions of the pseudo-2D simulation in the nominal benchmark case

As mentioned in the beginning of Section 3, we present here the results of single- and double-region pseudo-2D axial-azimuthal (x-y) simulations in terms of the time-averaged axial plasma profiles for the nominal benchmark case.

Before comparing the results with the benchmark, however, we refer to Figure 3 where we have compared the distribution of the plasma potential from our pseudo-2D simulations with that from a full-1D PIC simulation we performed without any ad-hoc collisionality (the dashed green line). In our full-1D simulation case, the axial electron transport is insufficient, and the electric potential is seen to not be sustained in ionization region near the left side of the domain. However, noting the plasma potential profiles from the single- and double-region simulations, it is clear that the pseudo-2D PIC scheme is capable of properly incorporating the necessary wave-induced mobility for the sustainment of the plasma potential. This verifies that our main goal behind the development of the pseudo-2D PIC is met; indeed, this scheme allows a code with the computational cost of a 1D simulation to self-consistently account for instability-induced electron transport.

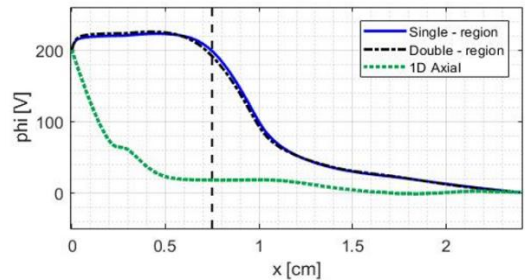

Figure 3: Axial profile of the plasma potential averaged over 16-18 μs from a 1D PIC simulation without ad-hoc electron mobility and the single- and double-region pseudo-2D simulations

Another remarkable aspect can be observed by referring to Figure 4, which presents the comparison with the benchmark of time-averaged plasma properties, electric field ($E_x$), electron temperature ($T_e$), and ion density ($n_i$). Profiles of these parameters from our pseudo-2D simulations are compared with the mean value of the corresponding profiles from the different PIC codes reported in [13]. It is seen that both the single-region and the double-region cases have provided results very similar to the 2D benchmark. Nevertheless, the results of the double-region implementation show a better agreement with the benchmark since this implementation provides the pseudo-2D simulation with the needed degree of freedom to resolve the transition in the waves characteristics [13].

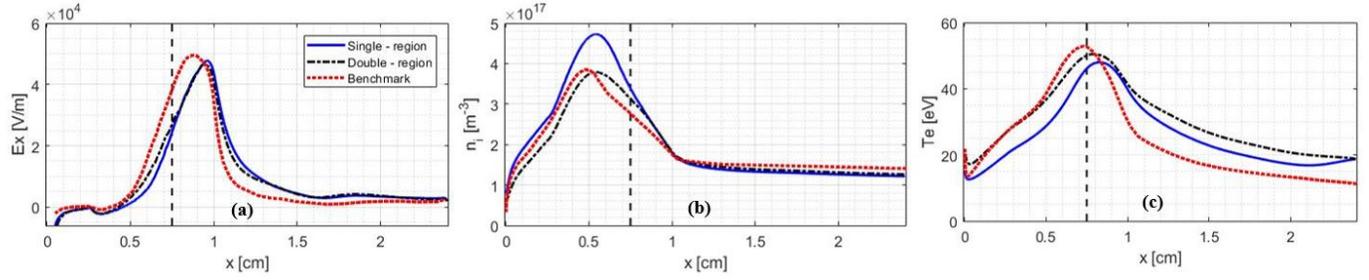

Figure 4: Axial profiles of plasma properties averaged over 16-18 μs: (a) $E_x$, (b) $n_i$, and (c) $T_e$. Black dashed lines show the location of the peak magnetic field.

In terms of the necessary computational resources, the major benefit of our pseudo-2D simulation comes from the fact that the total number of cells and, therefore, the total number of required macroparticles, is reduced from order $N^2$ (corresponding to a full-2D simulation) to order $N$, where $N$ is the number of nodes. In this respect and referring to the nominal case discussed in this section, the conventional 2D kinetic codes that participated in the benchmarking activity of Ref. [13] required about 2.5 to 11 days for a complete run when using about 100 to 360 CPU cores. This is whereas our simulation took 3 days using only a single CPU core on a standard personal workstation.

### Section 4: Pseudo-2D axial-azimuthal simulations for a range of operating conditions

We performed pseudo-2D simulations over a range of operating conditions to further assess the predictive capability of the approach. In the following, the results of the single- and double-region simulations with various current densities (representing different anode mass flow rates) are presented and compared against the full-2D results published in [28]. The simulation conditions and setup are similar to those reported in Section 3.1, and only the peak value of the ionization source is adjusted for different current densities.

Figure 5 and Figure 6 show the profiles of the time-averaged plasma properties. Overall, the axial profiles from both the single- and double-region simulations agree well at all current levels with those from the 2D simulations. However, the double-region simulations indicate a slight improvement in the profiles, which is evident from the plasma potential (Figure 5, bottom row) and the ion density profiles (Figure 6, first row).

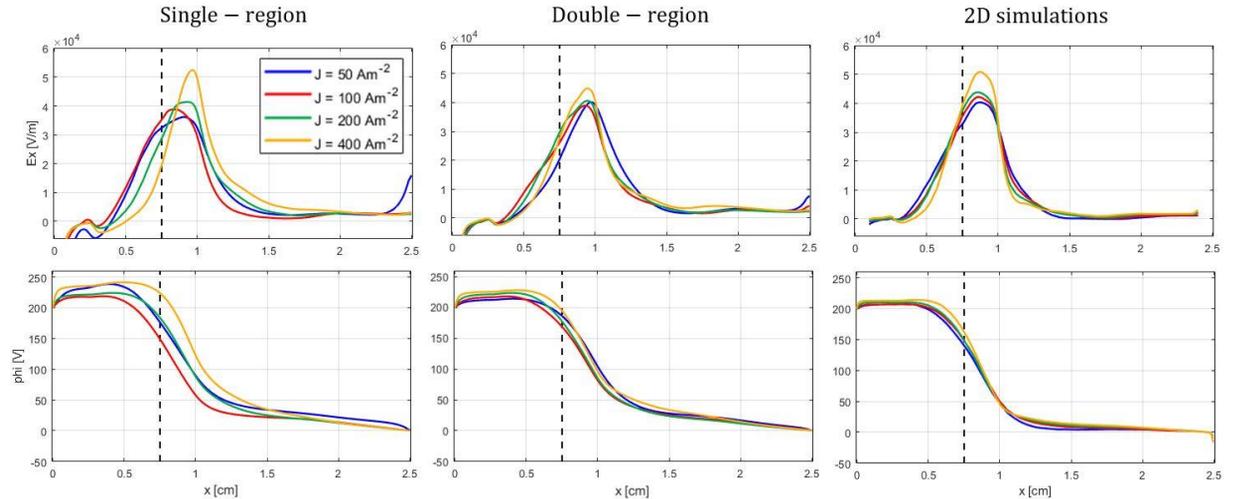

Figure 5: Time-averaged axial profiles of axial electric field (first row) and plasma potential (second row) for various current densities from the single- and double-region approach, and from the 2D simulations of [28].

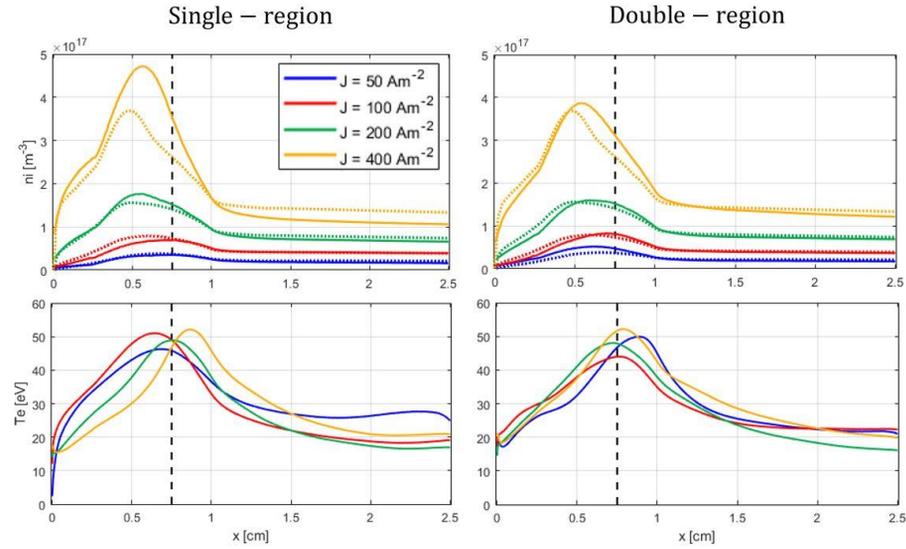

Figure 6: Time-averaged axial profiles of ion number density (first row) and electron temperature (second row) for various current densities from the single- and double-region approach (solid lines) against the 2D results of [28] (dashed lines)

As it was mentioned in Section 2.1, the two vertical regions in the double-region simulations are selected such that the characteristics of the waves would remain almost constant in each of them. Indeed, the transition from short- to long-wavelength oscillations is observed in 2D PIC simulations [13][28] to occur near the ion sonic point that is located within the region of peak magnetic field. For the simulation results in this section, the boundary between the two regions was set at the location of the peak magnetic field intensity ($x = 0.75\ cm$). In this regard, to ascertain that this choice of the boundary location is consistent with the ion sonic point, we plotted in Figure 7 the axial distribution of the ion Mach number for various current densities. It is observed that, across all cases, the ion sonic point occurs within the range of $x \in [0.75, 0.83]$ cm, which is close to our assumption. It is also noteworthy that the ion sonic point shows a smaller variation with respect to current density in the double-region simulations, an observation that is consistent with the corresponding plasma potential plot in Figure 5.

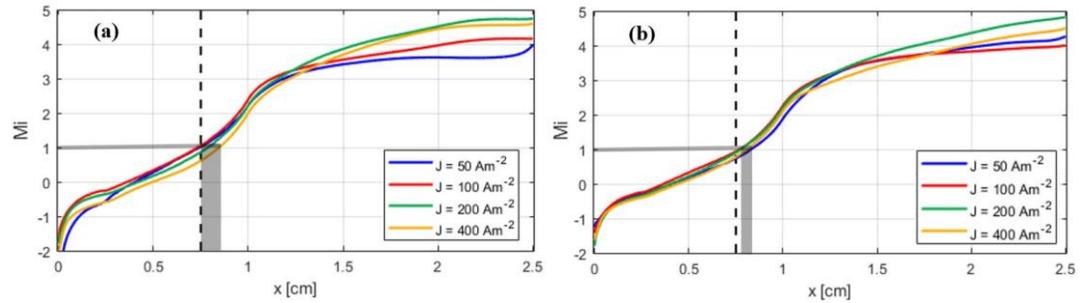

Figure 7: Axial profile of the ion Mach number in (a) single-region and (b) double-region simulations for different current densities. The grey region represents the axial positions of the ion sonic point for various J values.

### 4.1. Characteristics of the azimuthal fluctuations

Previous numerical and experimental works [12] have demonstrated that, along the azimuthal direction of a Hall thruster, the short-wavelength, high-frequency Electron Drift Instability (EDI) waves become dominant near the thruster's channel exit and in the near-plume region due to a rapid non-linear growth. The saturation of these modes through ion trapping, is observed to be accompanied by a transition into longer-wavelength, lower frequency waves representative of Ion Acoustic Instability (IAI) [29]. Both instabilities are shown to have a significant contribution to enhancing electron cross-field transport [12].

In our case, when the simulations arrive at steady state, which takes about 10 $\mu s$, the observed azimuthal oscillations are expected to have similar characteristics to the IAI modes. Consequently, an important verification step concerning the pseudo-2D PIC scheme is to ensure that the underlying azimuthal physics and the resulting enhanced electron mobility is being properly resolved. To this end, we analyze in this section the captured azimuthal electric field fluctuations and their dispersion characteristics, comparing against the theoretical dispersion relation of the modified Ion Acoustic Instability.

In Figure 8, the single- and double-region simulation results in terms of the fluctuations in the azimuthal electric field for various current densities are shown. Referring to the plots from the double-region simulations, the transition in the wave characteristics is clearly observable: In region I, the waves have higher frequency and shorter wavelength (larger wavenumber), whereas in region II, the waves are of lower frequency and longer wavelengths (smaller wavenumber). Due to the lack of distinction between the two regions in the single-region simulations, the waves' characteristics and dynamics show an average representation of the corresponding properties and features from the double-region simulations. To better illustrate this point, and to also compare the oscillations' characteristics with those of the IAI modes, we applied the 2D FFT to the azimuthal electric field signals in Figure 8 to obtain the dispersion map of the instabilities in the $(k_y - \omega_R)$ plane, with $k_y$ being the azimuthal wave number and $\omega$ the real frequency component. The results are shown in Figure 9, in which the amplitudes are normalized with respect to that of the dominant mode. The same normalization is performed in all other dispersion plots presented in this paper.

Superimposed on the dispersion maps are the plots of the theoretical dispersion relation of the ion acoustic waves in the laboratory frame [30]

$$\omega_R \approx \vec{k} \cdot \vec{V}_{di} \pm \frac{k_y C_s}{\sqrt{1 + k_y^2 \lambda_D^2}}, \tag{Eq. 8}$$

$$\gamma \approx \pm \sqrt{\pi \frac{m_e}{m_i}} \frac{\vec{k} \cdot \vec{V}_{di}}{\left(1 + k_y^2 \lambda_D^2\right)^{\frac{3}{2}}}, \tag{Eq. 9}$$

where $\vec{k}$ is the wavenumber vector, $\vec{V}_{di}$ is the ions' drift velocity, $C_s$ is the ion sound speed, $\lambda_D$ is the Debye length and $\gamma$ is the growth rate. It should be noted that, in our case, $\vec{V}_{di}$ is mostly in the axial direction and, as the wavenumber is mostly in the azimuthal direction, we have $\vec{k} \cdot \vec{V}_{di} = 0$.

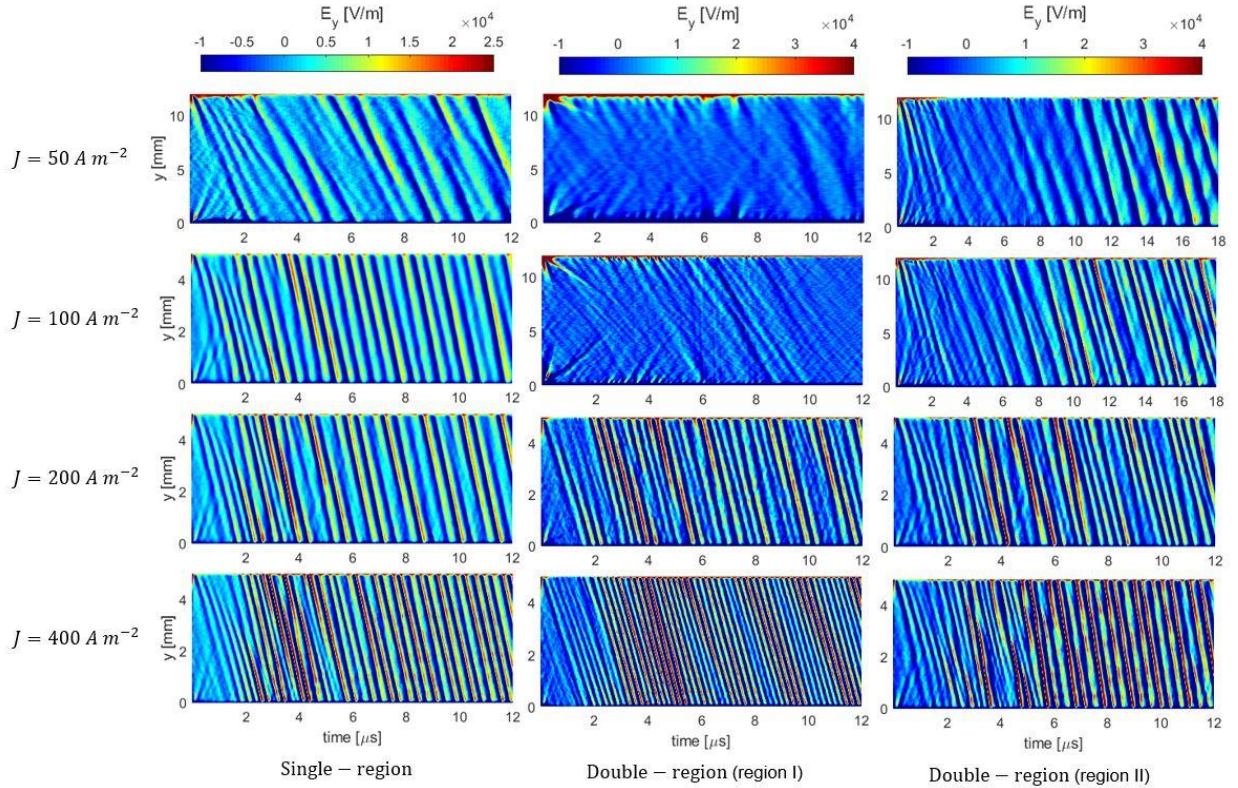

Figure 8: Azimuthal electric field fluctuations from the single-region (first column), and region I (second column) and region II (third column) of the double-region simulations for different current densities

Overall, the instabilities dispersion maps in Figure 9 are quite reasonably in agreement with the dispersion relation of the ion acoustic waves. Furthermore, the transition in waves' characteristics from region I to region II of the double-region simulations and the average representation captured in the single-region simulations, observed in the plots of Figure 8, are also reflected in these maps. In addition, focusing on the FFT plots related to the "nominal" benchmark case with $J = 400\ A.\,m^{-2}$ (the bottom row in Figure 9), we noticed a remarkable agreement when we compared these results with those published in [13].

The deviation of the resolved dispersion characteristics in region II of the double-region simulations (Figure 9) from the theoretical ion acoustic dispersion relation, Eqs. 8 and 9, is also observed by other authors (such as in [28]) and is attributed to the downstream convection of the instability waves. This demonstrates that our pseudo-2D approach is able to resolve this important feature concerning the azimuthal instabilities behavior in Hall thrusters. In this regard, since the plasma macroparticles are "shared" between the in-parallel axial and azimuthal simulations, the capability to capture the waves' convection, at least to some extent, was indeed expected.

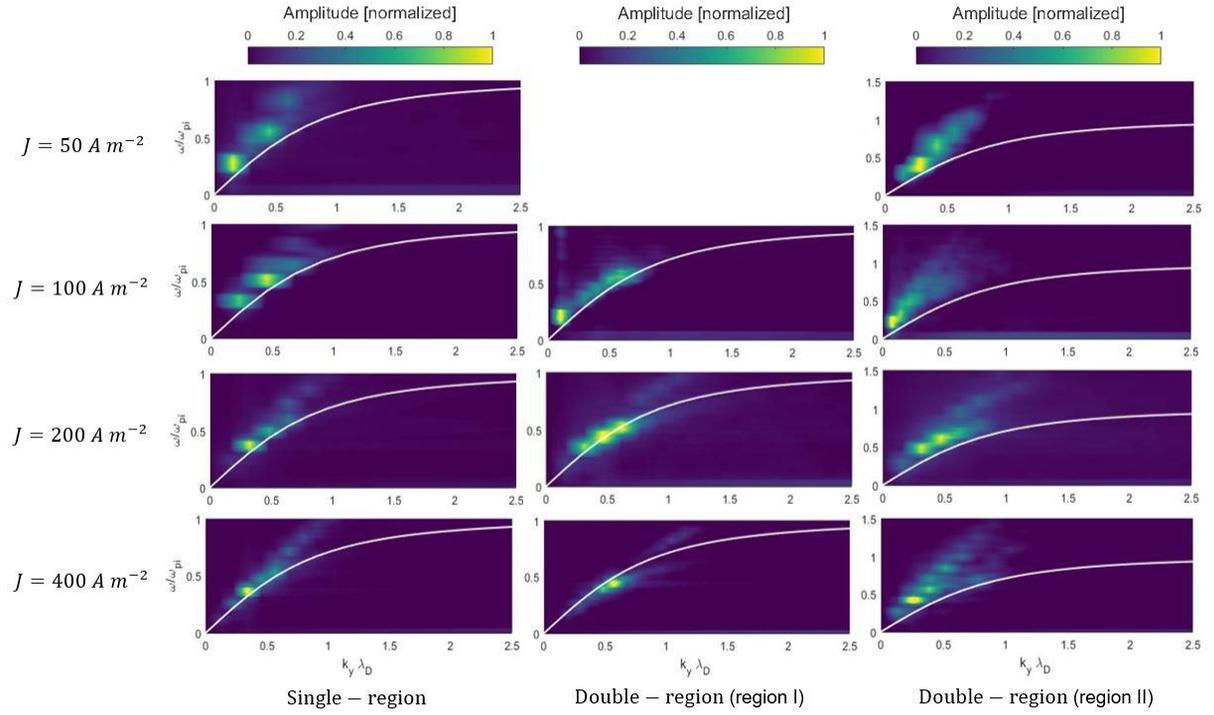

Figure 9: Dispersion plots of the azimuthal electric field oscillations observed in the single- and the double-region simulations for different current densities; the frequency ($\omega$) and azimuthal wavenumber ($k_y$) are normalized with respect to the local ion plasma frequency and the Debye length, respectively. For $J = 50\ A.\,m^{-2}$, the FFT plot for region I of the double-region simulation is not shown due to the lack of excitation of any coherent azimuthal waves in this region.

As the final point, looking back at Figure 8, we see that the amplitude and the frequency of the waves increase with the current density (or equivalently, the plasma number density). The observed correlation between the real frequency and the plasma density is, in particular, consistent with the theoretical dispersion relation of the IAI waves (Eq. 8) because, as the plasma density increases, so too does the real component of the waves' frequency. The reason behind the increase in amplitude with the plasma density is not studied in detail here but is speculated to be due to stronger wave-particle interactions at higher plasma densities during the EDI growth phase.

### 4.2. Contribution of force terms in the electron azimuthal momentum equation to cross-field transport in the presence of the azimuthal instabilities

To assess the fidelity of the results in terms of enhanced electron mobility for various current densities, we evaluated the contributions of the different terms in the electron azimuthal momentum equation to the cross-field transport. In this regard, we pursued an approach similar to that described in [11]. We have taken the velocity moment of the Vlasov's equation to derive the electron momentum equation which, along the azimuthal direction $z$, is written as

$$-qn_e v_{e,x} B = \partial_t \big( m n_e v_{e,z} \big) + \partial_x \big( m n_e v_{e,x} v_{e,z} \big) + \partial_x \big( \Pi_{e,xz} \big) - q n_e E_z \qquad \text{(Eq. 10)}$$

In the above equation, $q$ is the elementary charge, $n_e$ is the electron number density, $v_{e,x}$ and $v_{e,z}$ are the electron axial and azimuthal drift velocity, $B$ is the magnetic field intensity, $m$ is the electron mass and $E_z$ is the azimuthal electric field. In Eq. 12, the left-hand-side term is called the magnetic force ($F_B$) whereas, on the right-hand-side (RHS), the first term is the temporal inertia ($F_t$), the second term is the convective inertia ($F_I$), the third term corresponds to the off-diagonal components of the pressure tensor ($F_{\Pi_{xz}}$), and the last term is the electric force term ($F_E$) comprising both the steady ($\bar{n}_e \bar{E}_z$) component and the fluctuating ($\delta n_e \delta E_z$) component. The steady component is, however, zero since $\bar{E}_z = 0$, whereas due to a non-zero correlation between the electron density and the wave electric field oscillations, the fluctuating component can contribute to the electrons axial transport.

The force terms introduced above are the moments of the distribution function with the following definitions

$$F_B = -q \int_{-\infty}^{\infty} v_x \, B \, f_e(v) d^3v \,, \qquad \text{(Eq. 11)}$$

$$F_t = m_e \frac{\partial}{\partial t} \left( \int_{-\infty}^{\infty} v_z \, f_e(v) d^3v \right), \qquad \text{(Eq. 12)}$$

$$F_I = m_e \frac{\partial}{\partial x} \left( \frac{\int_{-\infty}^{\infty} v_x f_e(v) d^3v \int_{-\infty}^{\infty} v_z f_e(v) d^3v}{\int_{-\infty}^{\infty} f_e(v) d^3v} \right), \qquad \text{(Eq. 13)}$$

$$F_{\pi_{xz}} = m_e \frac{\partial}{\partial x} \left( \int_{-\infty}^{\infty} v_x \, v_z \, f_e(v) d^3v - \frac{\int_{-\infty}^{\infty} v_x f_e(v) d^3v \int_{-\infty}^{\infty} v_z f_e(v) d^3v}{\int_{-\infty}^{\infty} f_e(v) d^3v} \right), \qquad \text{(Eq. 14)}$$

$$F_E = -q \int_{-\infty}^{\infty} E_z f_e(v) d^3v \,. \qquad \text{(Eq. 15)}$$

Figure 10 shows the axial distributions of the transport terms in Eq. 10, obtained from the single-region and double-region simulations and averaged over the last $1 \, \mu s$ of the simulations time. The $F_t$ contribution was found to be almost zero and, thus, its distribution is not displayed. We compared our profiles with those reported in Ref. [13] (for the nominal case) and in Ref. [28] (for other current densities) and found a high degree of similarity with the full-2D results. This serves as additional evidence that the primary aim of the pseudo-2D scheme, i.e., to self-consistently capture the axial electron transport, is successfully met. Indeed, the magnetic force term distribution for all values of current density, from both the single- and double-region simulations, was found to be closely in line with the 2D profiles [13][28], with the results from the double-region implementation demonstrating even a higher consistency. In addition, the electric force term ($F_E$) representing the instability-induced electron mobility has a significant contribution to the electron axial cross-field motion in all cases, especially in the plume where the transport is almost entirely due to the instabilities. This observation is perfectly in agreement with the outcomes of previous research activities regarding the wave-induced electron cross-field transport in Hall thrusters [12].

Another important feature to point out in the plots of Figure 10 is that the sum of all RHS terms in Eq. 10, shown by a dashed black curve, is perfectly equal to the magnetic force term (the left-hand-side of Eq. 10), demonstrating the satisfaction of the Vlasov's equation. This underlines the absence of any notable numerical effects that could have influenced the obtained results.

As a final remark concerning these profiles, a relative "symmetry" can be noticed between the viscous term ($F_{\Pi_{xz}}$) and either the convective ($F_I$) or the instability-related ($F_E$) force terms, which is an interesting observation that merits further investigation. However, it is important to highlight that the assumption of a prescribed ionization source may alter the physics of the problem and, consequently, affect the transport terms. Hence, a deeper assessment needs to be performed in simulations with self-consistent ionization. This is left for future work.

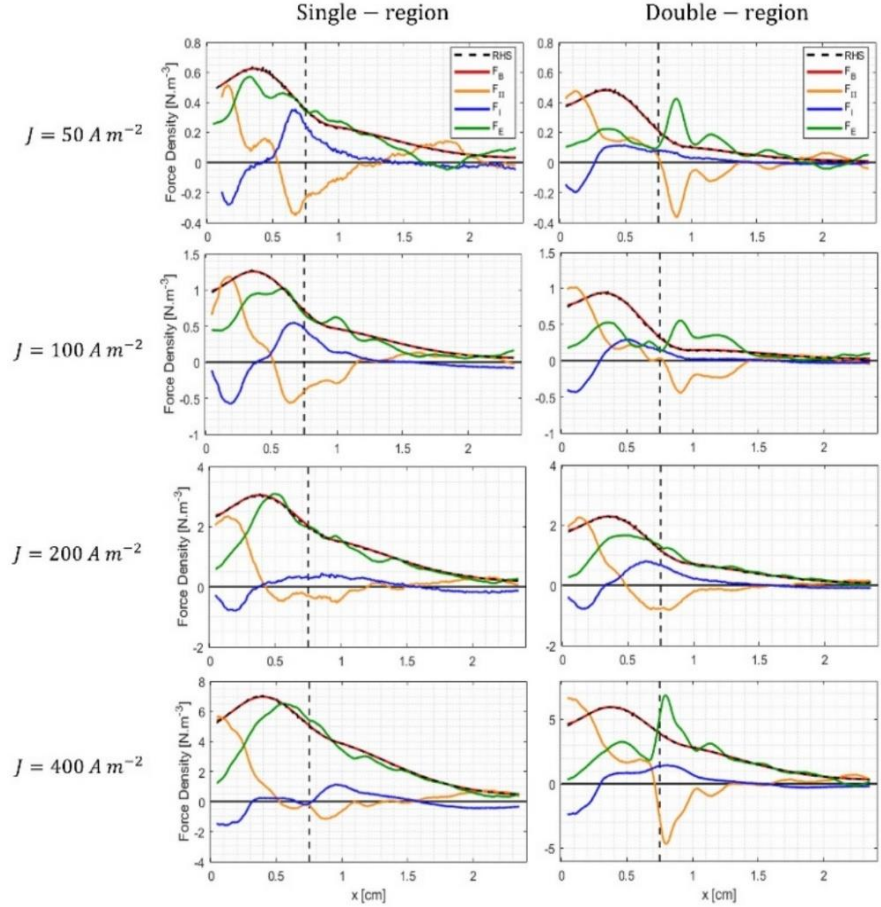

Figure 10: Axial profiles of the terms in Eq. 10, obtained from the single- and the double-region simulations for different current densities. The results in the last row correspond to the nominal benchmark case. The momentum terms are averaged over the last $1\mu s$ of the simulation time.

## Section 5: Analysis of the sensitivity to numerical and setup parameters

A sensitivity analysis was performed to evaluate the possible impacts that the choices of some numerical and setup parameters might have on the results. These parameters include the azimuthal length of the domain, the values of the weighting coefficients in Eq. 1, and the location of the boundary between the two vertical regions in the double-region simulation.

### 5.1. Sensitivity to the domain's azimuthal extent

As mentioned in Section 3.1, a periodic boundary condition is applied on the domain boundaries in the azimuthal direction. This boundary condition may have an influence on the resolved wave characteristics [25]. Consequently, using the single-region implementation, we repeated the pseudo-2D simulations of the nominal benchmark case for the azimuthal extents ($L_y$) of 1 and 2 cm and compared the resulting profiles of the time-averaged plasma properties and the azimuthal electric field fluctuations with those obtained from the simulation with $L_y = 0.5$ cm.

Figure 11 shows the influence of the azimuthal length on the plasma properties profiles. In addition, Figure 12 presents the spatio-temporal maps of the azimuthal electric field for the three simulated extents. Referring to Figure 11, the predictions of the three simulations are very similar to each other and are all consistent with the 2D benchmark results. Indeed, looking at Figure 12, a regular azimuthal wave pattern is seen to be formed, even with $L_y = 0.5$ cm, as the domain extent has been large enough with respect to the instabilities' wavelength. Indeed, in all our simulations, we observed that, when this was the case, the time-averaged plasma properties at steady state showed an almost negligible sensitivity to the azimuthal domain length, with only a slight increase in the electron temperature profile (Figure 11(d)) near the end of the domain due to a minor decrease in the waves' amplitude.

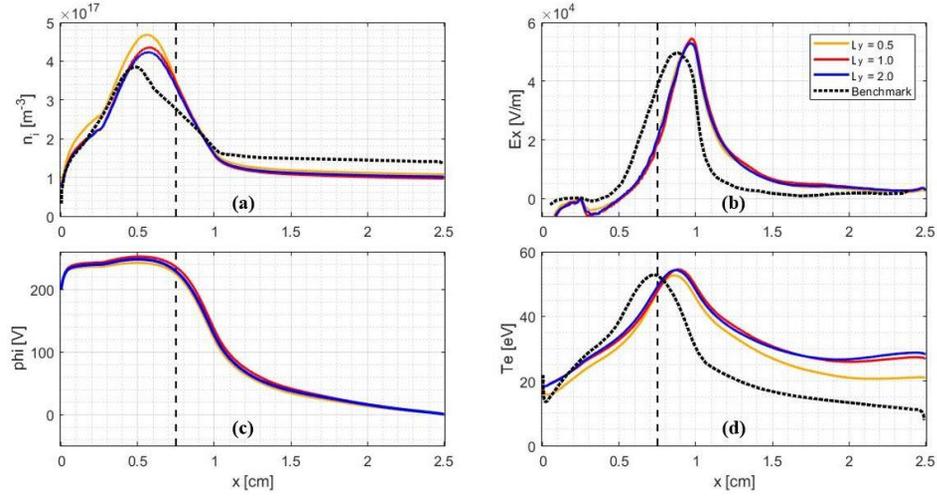

Figure 11: Comparison against the benchmark results [13] of time-averaged axial profiles, (a) ion number density, (b) axial electric field, (c) electric potential and (d) electron temperature, obtained from the single-region simulations with various $L_y$

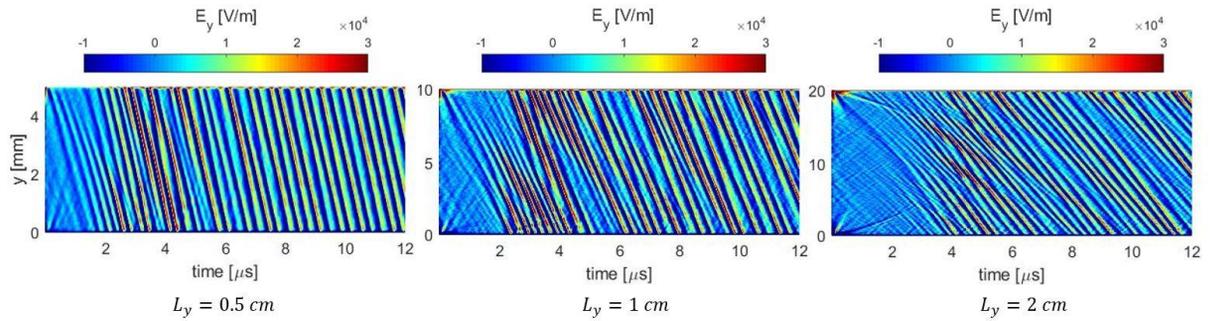

Figure 12: Azimuthal electric field fluctuations from the single-region simulations with various $L_y$

### 5.2. Sensitivity to the weighting coefficients in approximating the 2D potential field

Since we did not attempt to provide a rigorous mathematical definition for the coefficients $C^x$ ($WF_x$) and $C_1^y = C_2^y$ ($WF_y$) in this work, it was necessary to assess the sensitivity of the simulations' result to these parameters. The values of the coefficients mentioned in Section 2.1, where the axial and azimuthal potential functions ($\phi^x$, $\phi_m^y$) are equally weighted, are considered as the baseline values (denoted by WF0). The single-region simulation results with four other sets of values (Table 2) are compared here against the results with WF0 set.

|          | WF0  | WF1  | WF2  | WF3  | WF4  |
|----------|------|------|------|------|------|
| $WF_x$   | 1/2  | 3/4  | 1/4  | 5/6  | 1/6  |
| $WF_y$   | 1/2  | 1/4  | 3/4  | 1/6  | 5/6  |

Table 2: The Poisson's equation weighting factors used for single-region simulations

Figure 13 illustrates the comparison in terms of the time-averaged plasma properties. From the plots in this figure, even though some degree of sensitivity to $WF_x$ and $WF_y$ is noticed, the deviation is bounded, and the simulations remain stable over a wide range of the coefficient values. The maximum disparity is observed in the peak values of the plasma density and the location of the electron temperature peak.

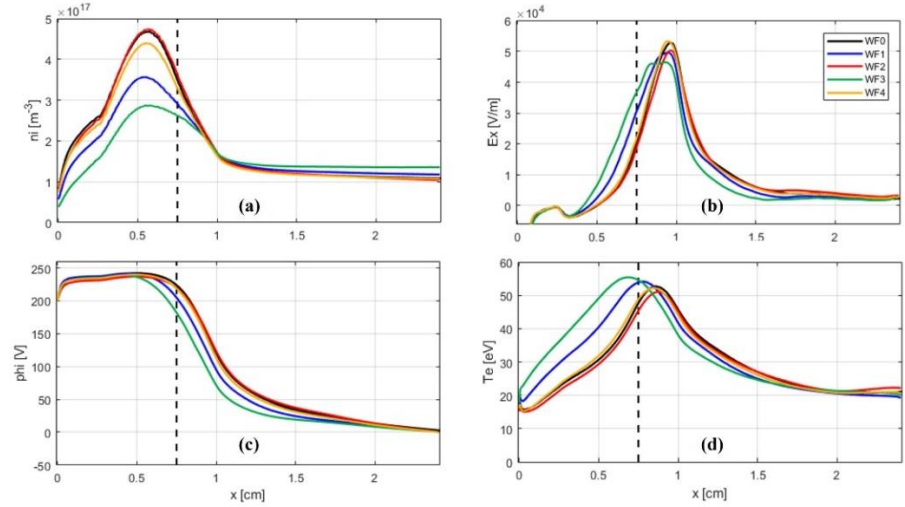

Figure 13: Comparison of time-averaged axial profiles of (a) ion number density, (b) axial electric field, (c) electric potential and (d) electron temperature, obtained from the single-region simulations with various Poisson's equation weighting coefficients

Aside from the intensive plasma properties profiles, to assess the effect of this sensitivity on the global performance parameters, we have shown in Table 3 the predictions of the thrust and thrust efficiency from the five simulations together with the corresponding errors with respect to the simulation with WF0. In this table, $I_e$, $I_i$ and $I_d$ are electron current, ion beam current and discharge current, respectively, which are obtained from the plasma profiles as

$$I_d = \frac{e\,A}{n\,N_{cell}} \sum_{ts=1}^{n} \sum_{j=1}^{N_{cell}} [n_e(j,t)|v_{ex}(j,t)| + n_i(j,t)|v_{ix}(j,t)|] \,, \qquad \text{(Eq. 16)}$$

$$I_i = \frac{e\,A}{n\,N_{cell}} \sum_{ts=1}^{n} \sum_{j=1}^{N_{cell}} n_i(j,t)\,|v_{ix}(j,t)| \,, \qquad \text{(Eq. 17)}$$

$$I_e = I_d - I_j \,. \qquad \text{(Eq. 18)}$$

In these equations, the inner summation denotes the average over the axial direction and the outer summation represents the average over the last 2 $\mu s$ of the simulation time. Thrust and efficiency are calculated using

$$T = \frac{A\,m_i}{n\,N_{cell}} \sum_{ts=1}^{n} \sum_{j=1}^{N_{cell}} n_i(j,t)\ v_{ix}^2(j,t) \,, \qquad \text{(Eq. 19)}$$

$$\eta = \frac{T^2}{2\dot{m}\,I_d\,V_d} \,, \qquad \text{(Eq. 20)}$$

where $V_d$ and $\dot{m}$ are, respectively, the discharge voltage and the anode mass flow rate. Eq. 19 neglects the less significant contributions to thrust from the ion and electron pressure terms, but it is sufficiently accurate for our intended comparative study here. To calculate the ion current and thrust, the average is taken over the last 100 cells of the domain where the ions were almost fully accelerated.

Comparing the performance parameters (Table 3), in all cases except for the case with $WF3$, the error in the thrust and thrust efficiency is less than 5%. This shows that, despite the sensitivity of the intensive plasma properties, the prediction of the performance is not notably affected by the selection of the weighting factors. Additionally, the simulations with $WF2$ and $WF4$, in which more weight is given to the azimuthal potential function, result in the least deviation from the reference profiles and performance.

Furthermore, looking at the instability maps and the corresponding 2D Fast-Fourier-Transform (FFT) plots (Figure 14), we can see that the above observation concerning the performance parameters is also true for the waves' characteristics. While the waves developed in simulations with $WF1$ and $WF3$ are of larger wavelength and smaller frequency, the waves in cases with $WF2$ and $WF4$ are rather similar to those in the reference case ($WF0$). This means that the simulation sensitivity is lower to the exact values of these coefficients as long as they are selected such that $WF_y > WF_x$.

| | WF0 | WF1 | WF2 | WF3 | WF4 |
|---|---|---|---|---|---|
| $I_e$ (A) | 1.33 | 1.46 | 1.11 | 1.48 | 1.25 |
| $I_i$ (A) | 1.24 | 4.32 | 1.20 | 1.45 | 1.22 |
| $I_d$ (A) | 2.57 | 2.78 | 2.3 | 2.93 | 2.47 |
| Thrust (mT) | 29.81 | 31.20 | 28.57 | 33.10 | 29.14 |
| $\eta$ (%) | 39.6 | 40.1 | 40.7 | 42.9 | 39.4 |
| $err_{Th}$ (%) | - | 4.66 | 4.16 | 11.04 | 2.25 |
| $err_\eta$ (%) | - | 1.26 | 2.77 | 8.33 | 0.5 |

Table 3: Performance parameters obtained from the single-region simulations with various Poisson's equation weighting coefficients

This analysis reinforces the applicability of the pseudo-2D scheme even in its current preliminary formulation where the precise definition of the weighting coefficients is not known. Nonetheless, we are currently developing a mathematically consistent formulation to provide an accurate approximation of the solution of the 2D Poisson's equation and to enable the deterministic calculation of the associated weighting coefficients. This effort is the central, first-priority part of the ongoing work on the subject.

### 5.3. Sensitivity to the location of the boundary between the two regions in the double-region simulations

As mentioned earlier, in the double-region simulations, the location of the boundary between the two regions ($x_b$) must coincide with the transition point of the instabilities where the characteristics of the waves change significantly. This transition location is observed to occur at about the ion sonic point [13][28].

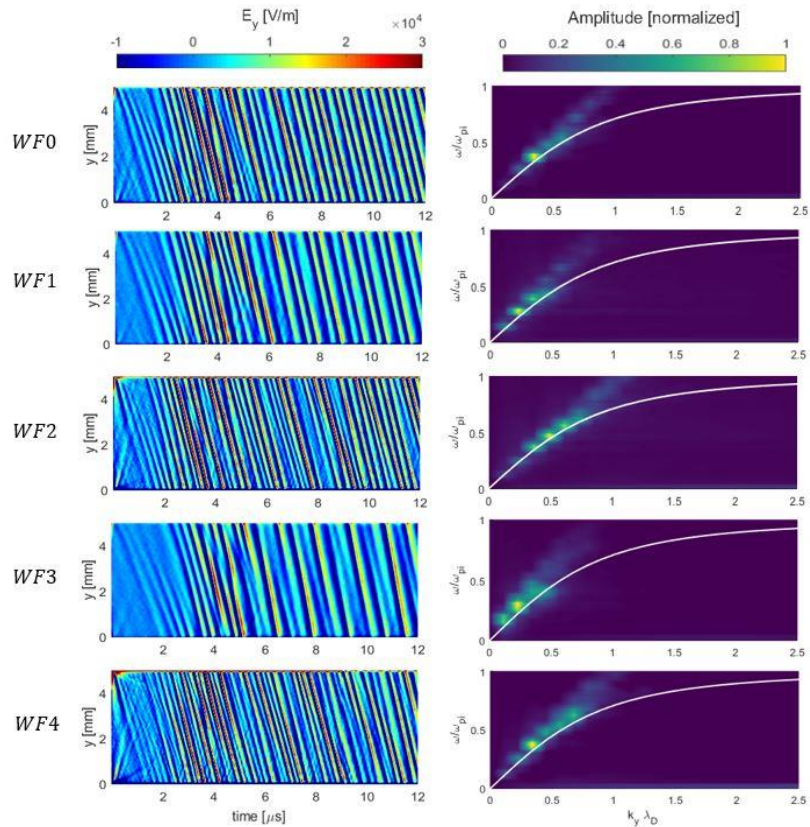

Figure 14: Azimuthal electric field fluctuations (left column) and the 2D FFT dispersion plots (right column), obtained from the single-region simulations with various Poisson's equation weighting coefficients

In this regard, to obviate the need for an a-priori knowledge of this transition point, we developed an algorithm that can update the location of the boundary through the simulation to where the ion Mach number becomes unity. However, for the double-region simulations whose results were presented Section 4, we prescribed the boundary at the location of the maximum magnetic field intensity ($x = 0.75\ cm$) for the entire simulation duration to be consistent with the full-2D benchmark. Indeed, in [13], it has been shown that, for the benchmark simulation case, the ion sonic location is very close to the point of maximum B-field. Nonetheless, noting that the boundary location is potentially a sensitive parameter, we assessed the influence of this parameter on the simulation results.

We carried out double-region simulations of the nominal benchmark case for two other boundary locations, namely, $x_b = 1\ cm$ and $x_b = 1.25\ cm$, and compared the change in the results with respect to the baseline simulation with $x_b = 0.75\ cm$. In Figure 15, we show the plots of the time-averaged plasma properties for the three different boundary locations and, for reference, also the full-2D benchmark results.

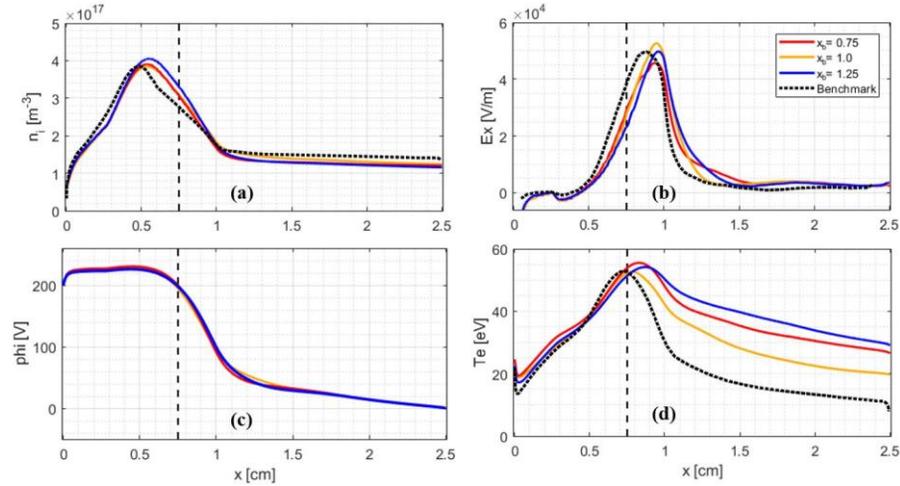

Figure 15: Comparison of time-averaged axial profiles of (a) ion number density, (b) axial electric field, (c) electric potential and (d) electron temperature, obtained from the double-region simulations with various locations of boundary between two regions ($x_b$), against the benchmark results [13]

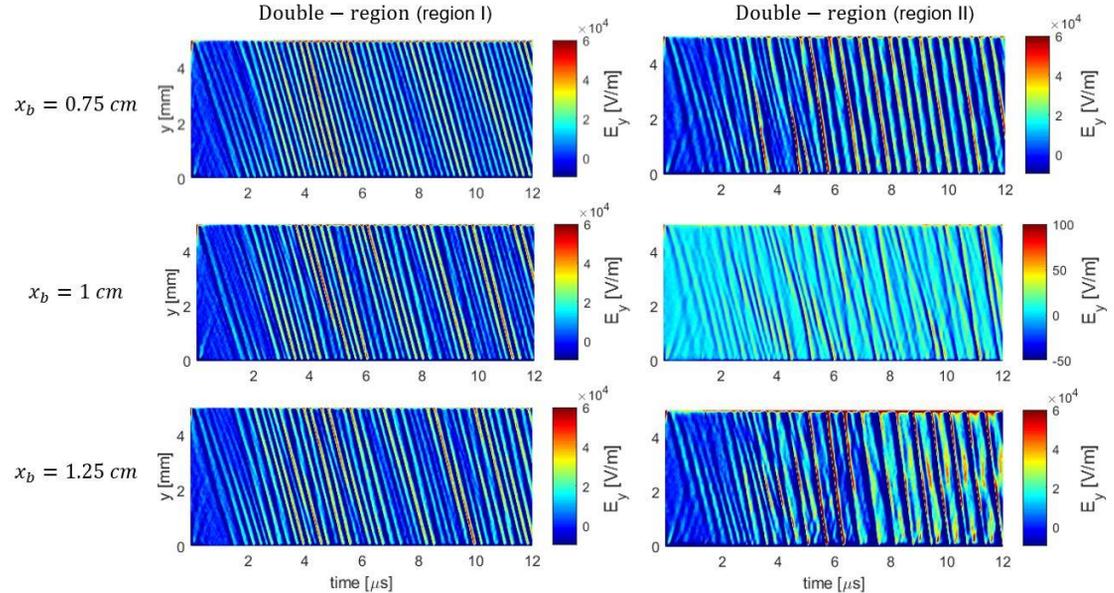

Figure 16: Azimuthal electric field fluctuations, obtained from the double-region simulations with various locations of the boundary between the two regions ($x_b$)

According to Figure 15, the axial profiles of the ion number density and the electric potential do not show any significant variation for different values of $x_b$ (0.75, 1 and 1.25 cm). Moreover, as expected for the double-region implementation, the results are in close agreement with those of the benchmark. However, the electron temperature in the near plume is affected by the value of $x_b$, which can be explained by noting the variation in the waves' intensity and evolution in region II (Figure 16). In this respect, it is recalled that the wave-particle interactions can lead to the heating of the bulk plasma and the thermalization of the electron population [3]. Hence, stronger wave-particle interactions when instabilities are of larger amplitudes can result in higher electron temperatures. Consequently, looking at the wave plots in Figure 16, it is observed that the waves in region I have almost the same intensities which lead to similar electron temperature profiles in this region. However, in region II, the waves' amplitudes are highest for the case where $x = 1.25$ cm which cause the highest electron temperatures. Similarly, for the case of $x = 1$ cm, the waves are orders of magnitude weaker in the plume leading to the lowest electron temperatures among the three cases.

Therefore, in view of these results, the adoption of an ion-sonic-point tracking algorithm for the self-consistent adjustment of the location of the boundary between the two regions is warranted.

### Section 6: Conclusions and future work

In the present article, we introduced an approach to approximate the 2D axial-azimuthal potential field in a Hall thruster that leads to a pseudo-2D description of the problem. The resulting pseudo-2D PIC scheme was shown to enhance the capabilities of a high-fidelity 1D PIC simulation to provide an accurate picture of the characteristics of the azimuthal instabilities and, thus, to self-consistently capture the instability-induced electron cross-field transport for multiple operating conditions.

We started by describing the IPPL PIC code that was the baseline for the pseudo-2D simulations. We discussed the double-region and the special-case single-region formulations underlying the pseudo-2D scheme. We highlighted that, in the current implementation, our method of approximating the 2D potential field in terms of a superimposition of 1D potential fields along the axial and azimuthal directions of the thruster yields a system of decoupled 1D Ordinary Differential Equations (ODE).

The results from the single- and double-region pseudo-2D simulations was presented for a range of operating conditions, consisting of multiple current densities (or equivalently, anode mass flow rates). We directly compared the time-averaged plasma properties profiles against those from the full-2D simulations of Ref. [13] and [28], and discussed the characteristics of the observed instabilities and their contribution to the electron transport. All results were found to be consistent with the reference 2D results. In particular, the similarity between the waves' characteristics from the pseudo-2D and the full-2D simulations shows that the deposition of the particles' charge on one or two azimuthal grids corresponding, respectively, to the single-region and the double-region case did not result in any significant artificial charge and/or electric field correlations, as described in [25], in the pseudo-2D simulations.

Additionally, the results of an extensive study on the simulations' parametric sensitivity were presented. We evaluated the influence of three parameters: the azimuthal domain length, the weighting coefficients to approximate the 2D potential field, and the location of the boundary between the two defined vertical regions of the double-region implementation. Concerning the azimuthal extent, it was observed that, despite slight differences in the waves' characteristics when increasing the azimuthal domain length, the sensitivity of the axial plasma profiles is negligible. It was, thus, concluded that to avert the sensitivity to the azimuthal length, it can be selected such that the formation of a regular wave pattern and the resolution of a few wavelengths of the instability will be guaranteed.

Regarding the weighting coefficients, we demonstrated that allocating more weight to the potential function along the azimuthal direction can minimize the sensitivity to this parameter in terms of either the time-averaged plasma properties or the waves' characteristics. Moreover, by selecting a larger weighting factor for the azimuthal potential, the error in the predicted performance parameters was also noted to be below 5%.

As for the sensitivity to the boundary location between the two regions, we highlighted the necessity for a proper selection of this boundary to coincide with the ion sonic point. In this respect, we pointed out that an algorithm is included in the code that can track the evolution of the ion Mach number through the simulation and to accordingly adjust the axial position of the boundary.

According to the performed analyses and studies, the remarkable results obtained underline the great promise of the pseudo-2D PIC scheme. However, the approach used to derive the pseudo-2D problem description is admittedly not complete in its current formulation and not readily extendable to other simulation configurations of interest for Hall thrusters such as the axial-radial one.

In this regard, we are working on some alternative formulations in order to develop a generalized "dimensionality-reduction" method for the Poisson's equation. This method is intended to allow increasing the number of regions arbitrarily along the different coordinates of the simulation domain for a general problem such that it would be possible to obtain an accurate approximation to the solution of the multi-dimensional Poisson's equation. Additionally, the alternative formulation is meant to enable deterministic calculation of any possibly involved weighting coefficients.

Implementing a generalized, mathematically consistent dimensionality-reduction method in a PIC code can lead to a "reduced-order" simulation scheme. Based on the fascinating potential observed for the pseudo-2D PIC scheme, the reduced-order PIC can retain the high-fidelity and self-consistency of the multi-dimensional Particle-in-Cell simulations while reducing the computational cost to effectively that of a 1D simulation by significantly lowering the required total number of macroparticles.

From an applied perspective, the reduced-order PIC scheme can minimize or eliminate the need for speed-up techniques currently in use for the simulations of full-scale high power Hall thrusters. Furthermore, the significant reduction in the computational cost compared to the traditional multi-dimensional PIC simulations can enable resolving the entire azimuthal extent of a Hall thruster such that the effects of non-uniformities along this direction, for example due to the propellant injection or cathode location, on the plasma behavior could be reliably studied. Ultimately, this PIC scheme can serve as a potential breakthrough towards achieving fully self-consistent pseudo-3D PIC simulations, thus providing a comprehensive picture of the plasma phenomena and the multi-dimensional effects in the thruster.

### Acknowledgments:

The present research is carried out within the framework of the project "Advanced Space Propulsion for Innovative Realization of space Exploration (ASPIRE)". ASPIRE has received funding from the European Union's Horizon 2020 Research and Innovation Programme under the Grant Agreement No. 101004366. The views expressed herein can in no way be taken as to reflect an official opinion of the Commission of the European Union.

### Data Availability Statement:

The data that support the findings of this study are available from the corresponding author upon reasonable request.

### Appendix:

#### A.1. Capacitively Coupled Discharge (CCP) Benchmark

The CCP benchmark [24] is one of the most rigorous in the low-temperature plasma community. It consists of a one-dimensional Cartesian domain enclosed between two planar, parallel electrodes driven by a sinusoidal AC voltage source, fluctuating at 13.56 MHz. The domain is filled with Helium gas at a given temperature and pressure (number density) that remain unvarying throughout the simulation. We have used the same collision cross-section data as in [24]. All physical and numerical parameters of the simulation have been set exactly as described in Table I of [24]. The ion number density has been suggested as the main parameter to use for comparison against benchmark results since it is highly sensitive to the implementation details, particularly in the MC collision module, and can be calculated in the most unambiguous manner. As such, in Figure 17, we have compared the ion number density profiles with respect to the published results in two cases, corresponding to Case I and Case II of [24]. It is observed that our code predictions are very closely aligned with the expected density distributions, and hence, we deem this verification as being successfully concluded.

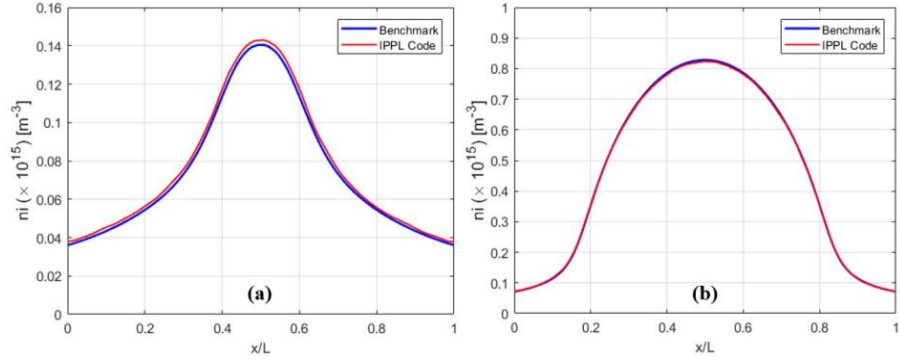

Figure 17: Comparison of ion density distributions obtained from the IPPL code with those of the benchmark [24]: (a) Case I, (b) Case II.

### A.2. 1D Azimuthal benchmark in Hall thruster representative conditions

We carried out another important benchmark to verify the capability of the code to accurately capture the fluctuations in plasma properties along the azimuthal coordinate of a Hall thruster. The waves associated with azimuthal oscillations in Hall thrusters are of relatively high-frequency (from $1-10$ MHz) and short wavelength (on the order of 1 mm) and exhibit non-linear behavior during the growth and saturation phases. As a result, a simulation aimed at investigating these waves can be prone to numerical effects that may alter their characteristics and/or their evolution dynamics.

Accordingly, we selected as the reference a simulation case [25] which has been used within the community by other authors as well for benchmarking [26][27]. The simulation domain consists of a 0.5 cm-long section of the azimuthal extent of a Hall thruster channel, with an infinitely long radial extent and a fictitious axial extent with the length of 1 cm. The problem description is essentially 1D as the Poisson's equation is only solved along the azimuthal ($y$) coordinate. The axial extent is included to provide a means to represent the axial convection of the instability waves, which is important to properly capture their saturation. In this regard, the electrons that leave the left axial boundary are resampled from their initial distribution and reloaded at the other end of the domain with a random azimuthal position. The same condition is applied to the ions that leave the right axial boundary of the domain. The rest of the setup and simulation conditions are identical to those adopted in [25].

Figure 18(a) and (b) show the saturated state of the azimuthal electric field and electron number density fluctuations normalized, respectively, with respect to the axial electric field and average plasma density in the domain. The observed characteristics of the waves (frequency, wavelength, and amplitude) were found to be very similar to the benchmark [25]. Moreover, the ion-wave trapping, also reported in the reference case, is evident from the ions phase-space distribution plot in Figure 18(c). This trapping of the ions in the waves potential well is shown to lead to the saturation of the azimuthal instabilities captured in this simulation. Finally, the total electron kinetic energy after the nonlinear saturation of the instability at about 0.5 $\mu s$ is seen to be around 70 eV, which is consistent with the values presented in [25].

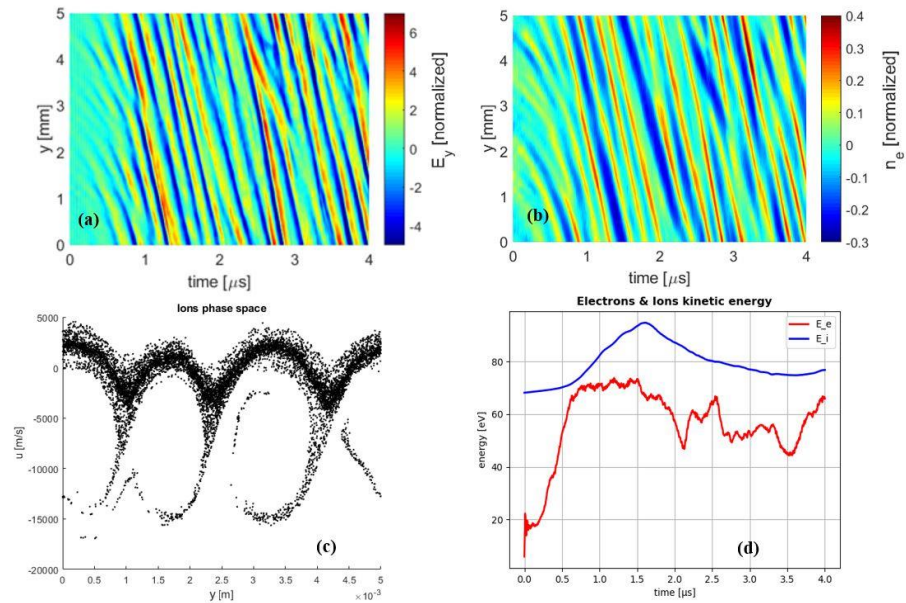

Figure 18: Results of our simulation of the 1D azimuthal benchmark case [25]: spatio-temporal maps of (a) normalized azimuthal electric field ($E_y$), and (b) normalized electron number density ($n_e$). (c) ion particles phase plot, (d) time evolution of the electrons and ions total kinetic energy

**References**:

[1]   Hargus Jr. W.A. and Nakles M.R. "Hall effect Thruster Ground Testing Challenges", *25th Aerospace Testing Seminar* (2009)

[2]   Polk JE, Brophy JR, "Life Qualification of Hall Thrusters by Analysis and Test", SP00547, *Space Propulsion Conference*, Seville, Spain (2018)

[3]   Boeuf JP, "Tutorial: Physics and modeling of Hall thrusters", *Journal of Applied Physics* **121** 011101 (2017)

[4]   Faraji F, Reza M, Andreussi T, "Modular Comprehensive Modeling of Plasma Behavior in Hall Thrusters", IEPC-2019-147, *36th International Electric Propulsion Conference*, Vienna (2019)

[5]   Taccogna F and Garriguez L, "Latest progress in Hall thrusters plasma modelling", *Reviews of Modern Plasma Physics*, Springer Singapore, 3 (**1**) (2019)

[6]   Reza M, Faraji F, Andreussi T, Andrenucci M "A Model for Turbulence-Induced Electron Transport in Hall Thrusters", IEPC-2017-367, *35th International Electric Propulsion Conference*, Atlanta, Georgia (2017)

[7]   Jorns B, "Predictive, data-driven model for the anomalous electron collision frequency in a Hall effect thruster", *Plasma Sources Sci. Technol.* **27** 104007 (2018)

[8]   Mikellides IG, Jorns B, Katz I, Ortega AL, "Hall2De Simulations with a First-principles Electron Transport Model Based on the Electron Cyclotron Drift Instability", *52nd AIAA/SAE/ASEE Joint Propulsion Conference*, Salt Lake City, Utah (2016) DOI: 10.2514/6.2016-4618

[9]   Ortega AL, Katz I, Chaplin VH, "Application of a first-principles anomalous transport model for electrons to multiple Hall thrusters and operating conditions", *Joint Propulsion Conference*, Cincinnati, Ohio (2018) DOI: 10.2514/6.2018-4903

[10]  Hara K, Yamashita Y, Tsikata S, et al., "New insights into electron transport due to azimuthal drift in a Hall effect thruster", IEPC-2019-691, *36th International Electric Propulsion Conference*, Vienna (2019)

[11]  Lafleur T and Chabert P, "The role of instability-enhanced friction on 'anomalous' electron and ion transport in Hall-effect thrusters", *Plasma Sources Science and Technology* **27** 015003 (2018)

[12]  Kaganovich I.D. et al, "Physics of E × B discharges relevant to plasma propulsion and similar devices", *Physics of Plasmas* **27**, 120601 (2020)

[13]    Charoy T et al., "2D axial-azimuthal particle-in-cell benchmark for low-temperature partially magnetized plasmas", *Plasma Sources Sci. Technol.* **28** 105010 (2019)

[14]    Vahedi V, Surendra M, "A Monte Carlo collision model for the particle-in-cell method: applications to argon and oxygen discharges", *Computer Physics Communications* **87** 179-198 (1995)

[15]    Alexander FJ, Garcia AL, "The Direct Simulation Monte Carlo method", *Computers in Physics* **11**, 588 (1997)

[16]    Croes V, "Modélisation bidimensionnelle de la décharge plasma dans un propulseur de Hall", PhD Dissertation (2018)

[17]    Vaughan J, "A New Formula for Secondary Emission Yield", *IEEE Trans. on Electron Devices* **36**, 9 (1989)

[18]    Bezanson J, Edelman A, Karpinski S, Shah VB, "Julia: A Fresh Approach to Numerical Computing", *SIAM Review*, **59**: 65–98 (2017)

[19]    Blackman D, Vigna S, "Scrambled linear pseudorandom number generators". *ACM Trans. Math. Softw.*, **47**:1−32 (2021)

[20]    Hestenes MR, Stiefel E, "Methods of Conjugate Gradients for Solving Linear Systems". *Journal of Research of the National Bureau of Standards*. **49** (6): 409 (1952)

[21]    Langdon AB, Birdsall CK, "Theory of Plasma Simulation Using Finite-Size Particles", *The Physics of Fluids* **13**, 2115 (1970)

[22]    Birdsall CK, Langdon AB, "Plasma Physics via Computer Simulation", CRC Press, 1991

[23]    Boris JP, "The acceleration calculation from a scalar potential", Plasma Physics Laboratory, Princeton University, MATT-152 (1970)

[24]    Turner M.M et al., "Simulation benchmarks for low-pressure plasmas: Capacitive discharges", *Physics of Plasmas* **20** 013507 (2013)

[25]    Lafleur T, Baalrud SD, Chabert P., "Theory for the anomalous electron transport in Hall effect thrusters. I. Insights from particle-in-cell simulations," *Phys. Plasmas* **23**, 053502 (2016)

[26]    Hara K and Cho S, "Radial-azimuthal particle-in-cell simulation of a Hall effect thruster," IEPC-2017-495, *35th International Electric Propulsion Conference*, Atlanta, GA, (2017)

[27]    Katz I, Chaplin VH, Ortega AL, "Particle-in-cell simulations of Hall thruster acceleration and near plume regions", *Phys. Plasmas* **25**, 123504 (2018)

[28]    Charoy T, "Numerical study of electron transport in Hall thrusters", Plasma Physics [physics.plasm-ph], Institut Polytechnique de Paris. English. ⟨NNT: 2020IPPAX046⟩ ⟨tel-02982367⟩ (2020)

[29]    Janhunen S, Smolyakov A, Sydorenko D et al. "Evolution of the electron cyclotron drift instability in two-dimensions", *Physics of Plasmas* **25**, 082308 (2018)

[30]    Charoy T, Bourdon A, Chabert P, Lafleur T, Tavant A, "Oscillation analysis in Hall thrusters with 2D (axial-azimuthal) Particle-In-Cell simulations", IEPC-2019-A487, *36th International Electric Propulsion Conference*, Vienna (2019)

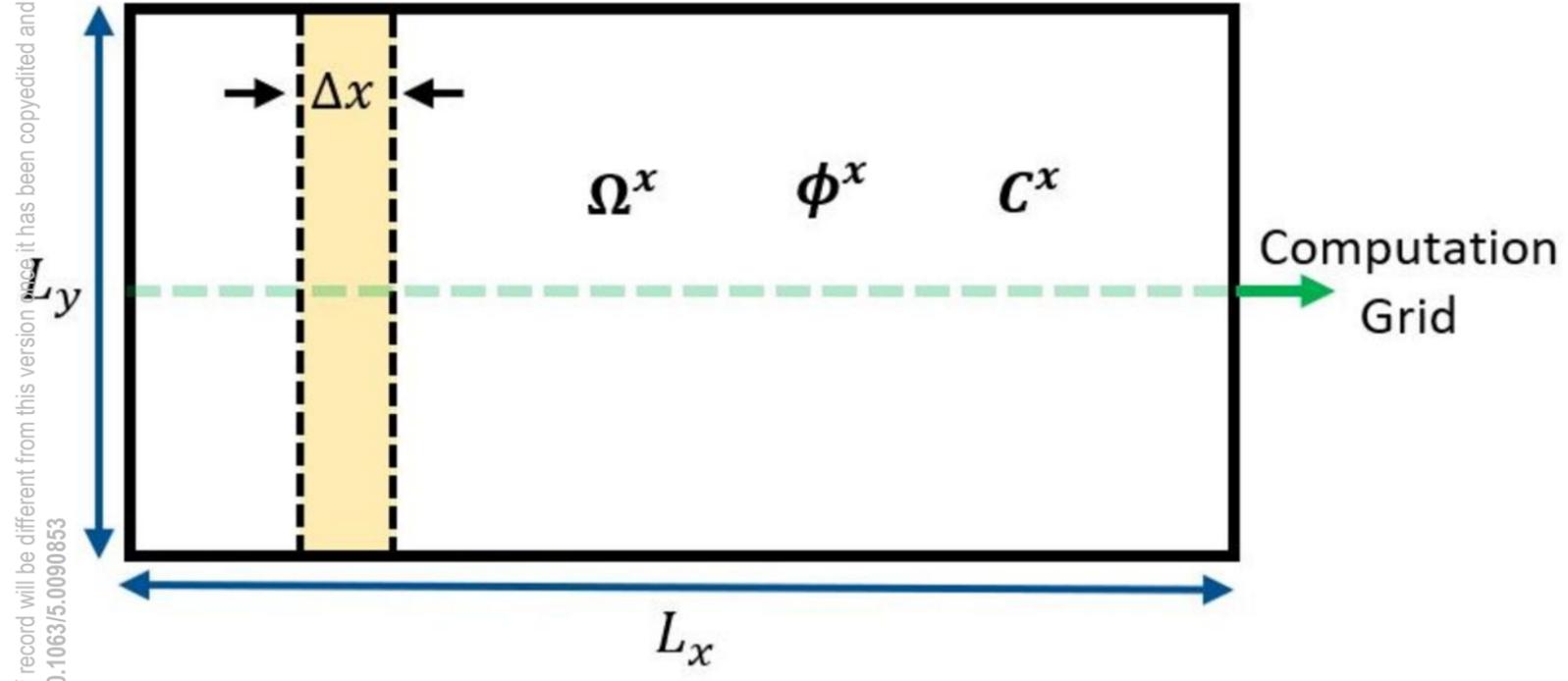

**(a)**

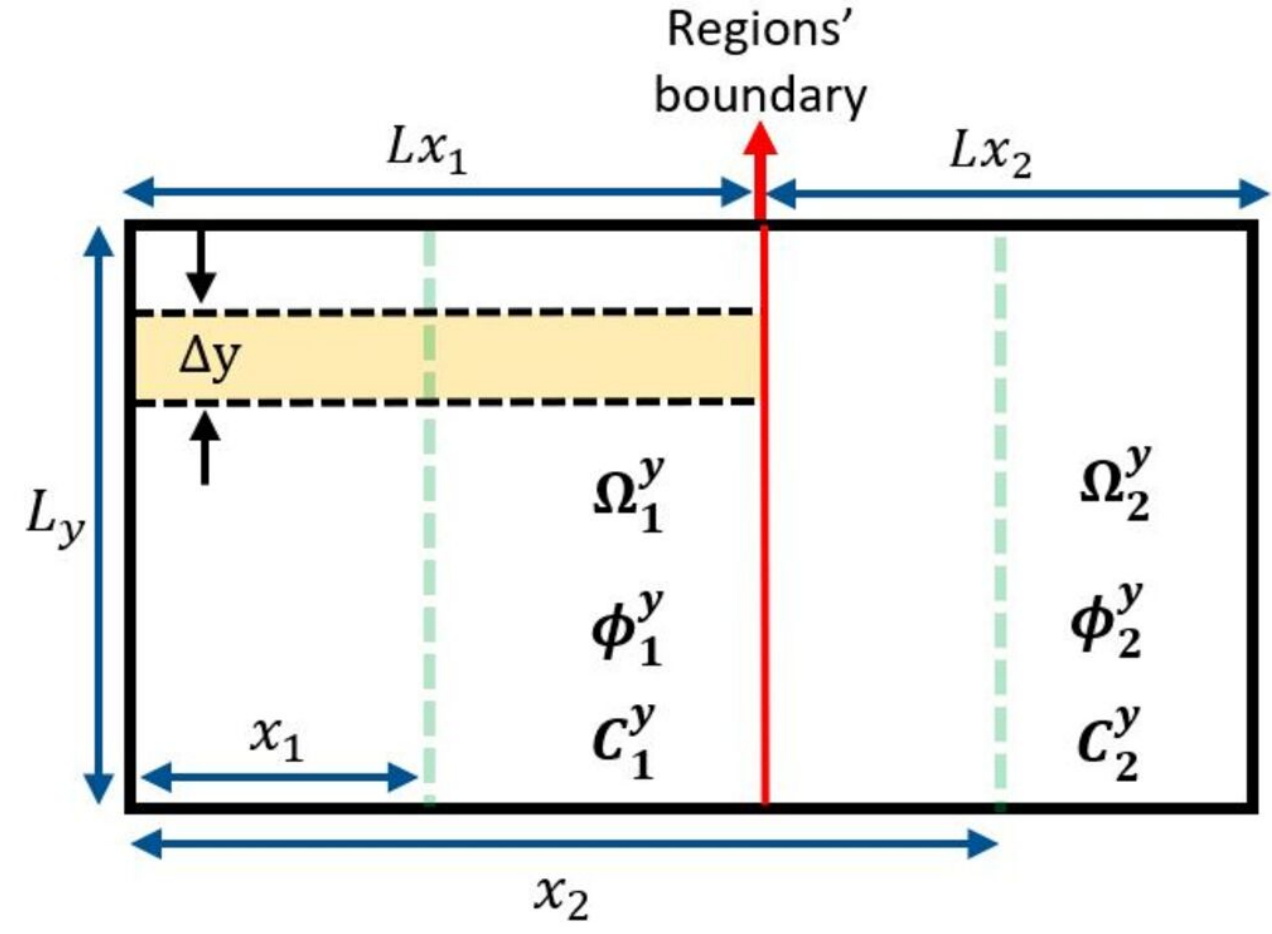

**(b)**

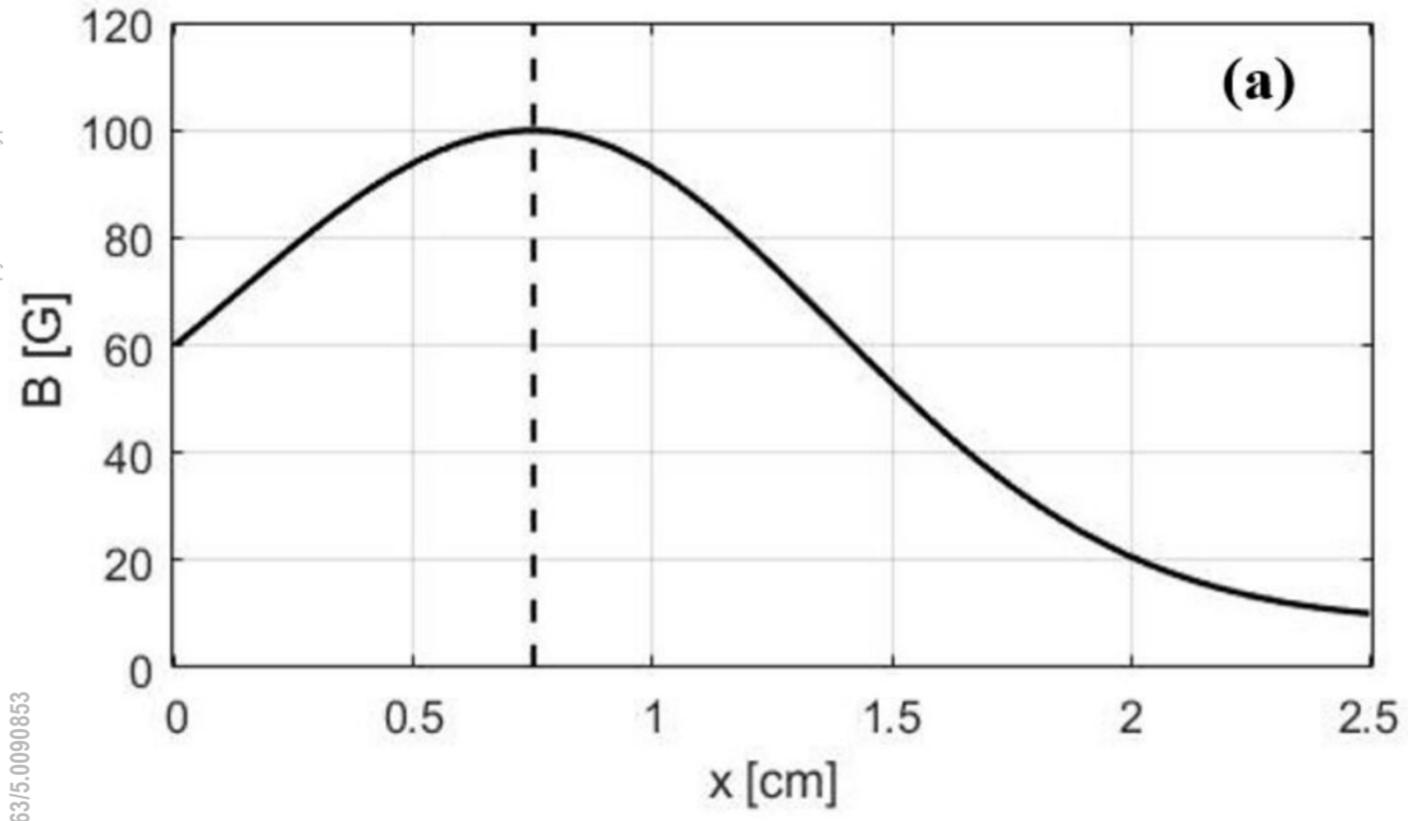

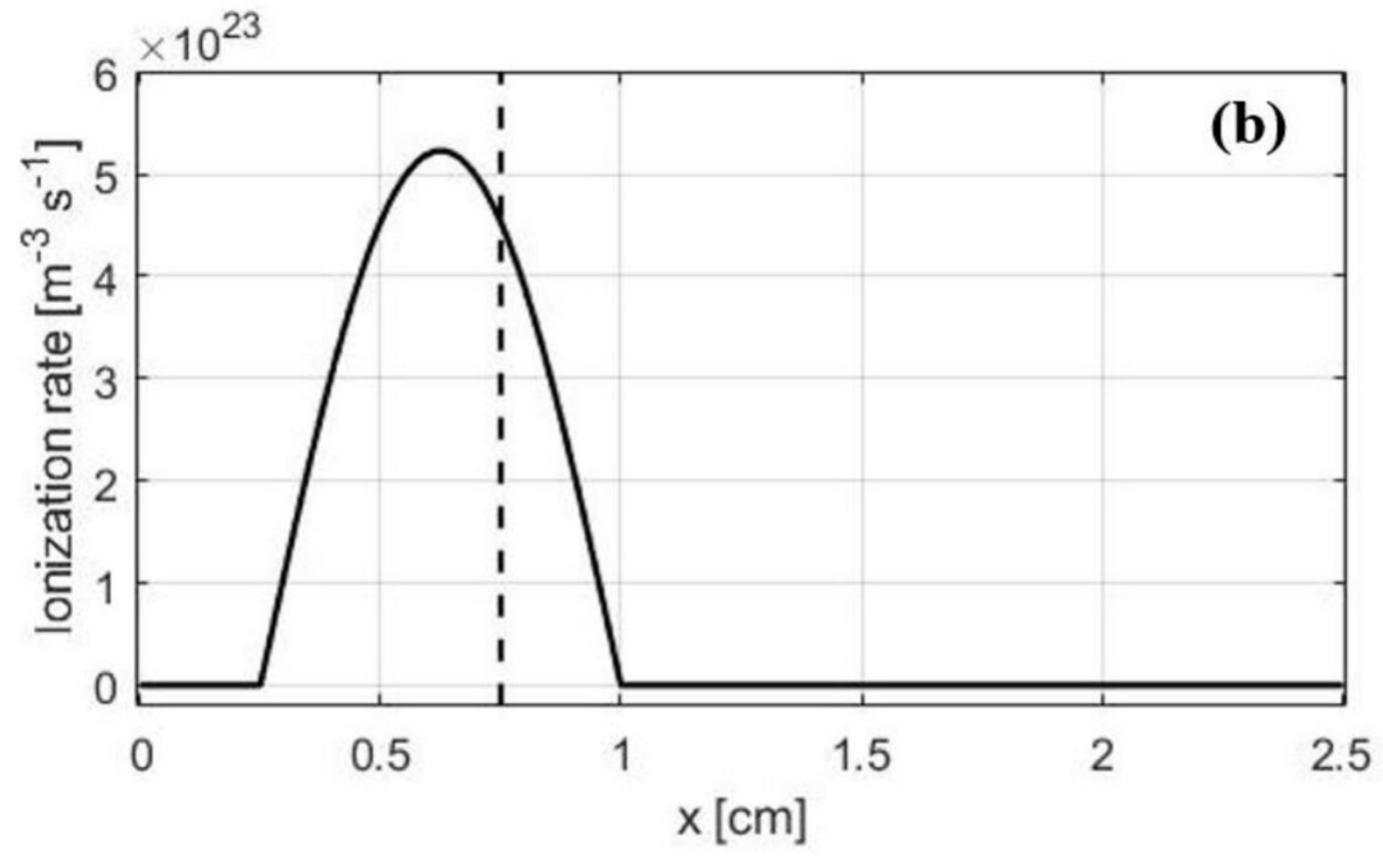

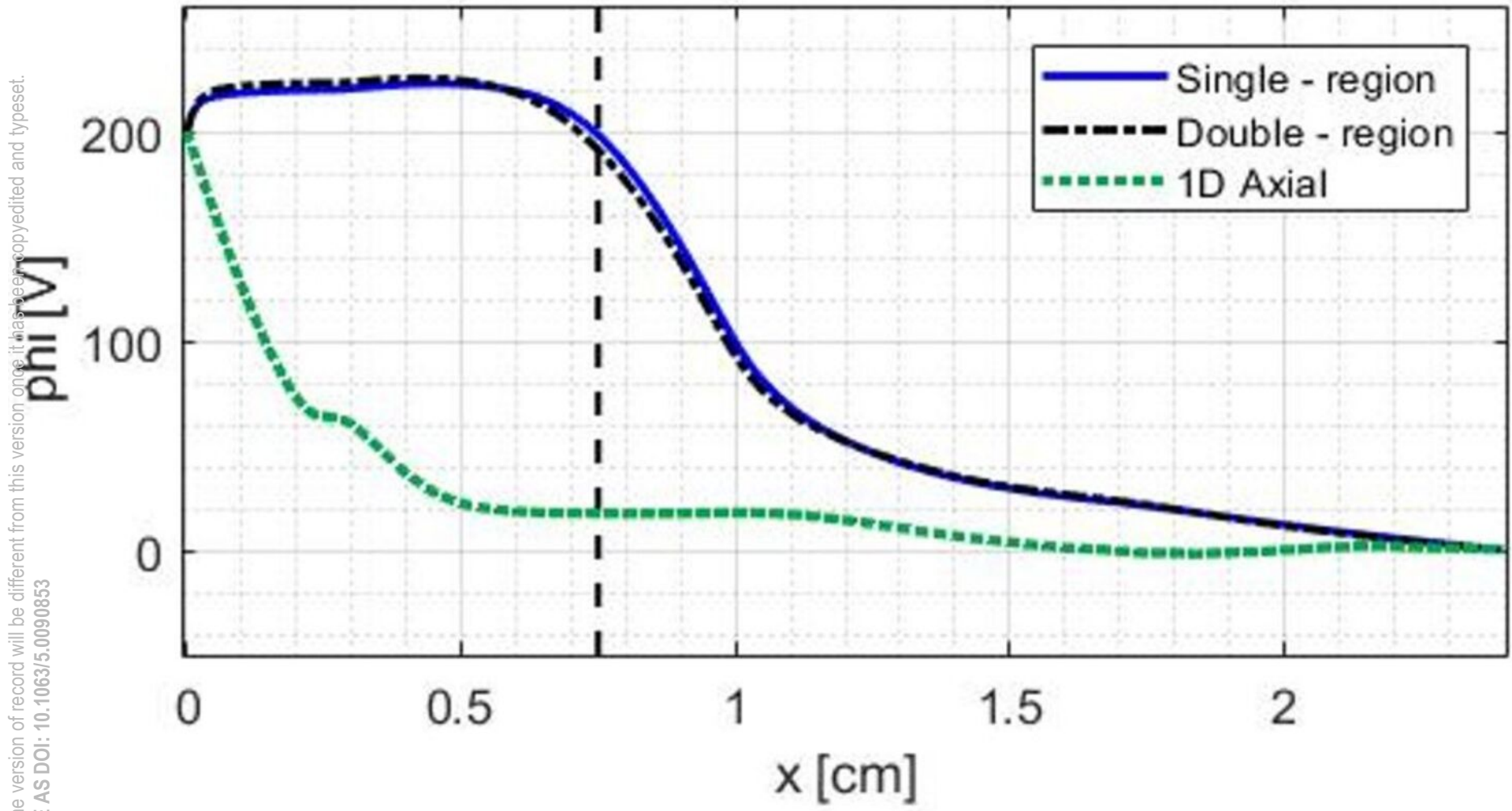

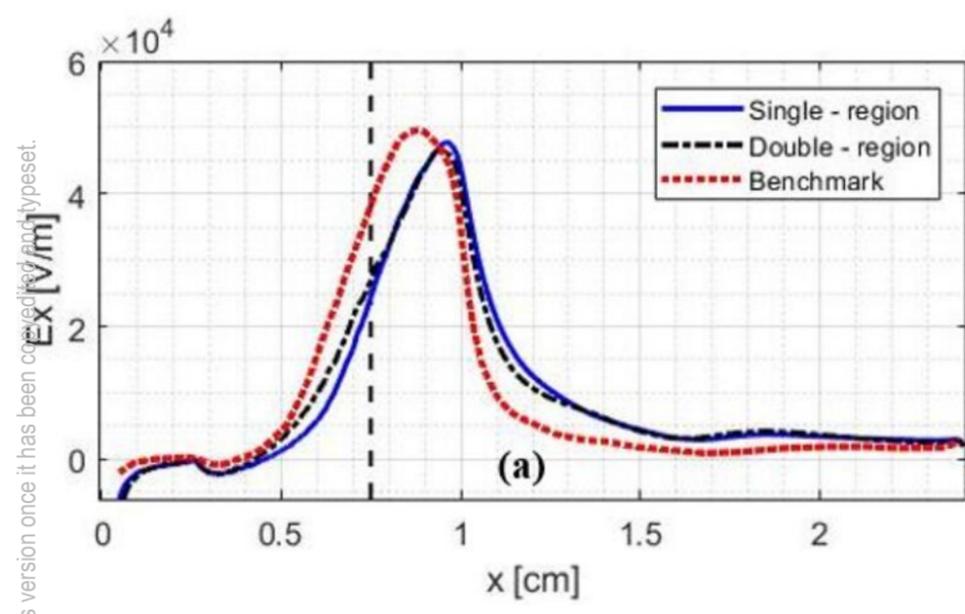

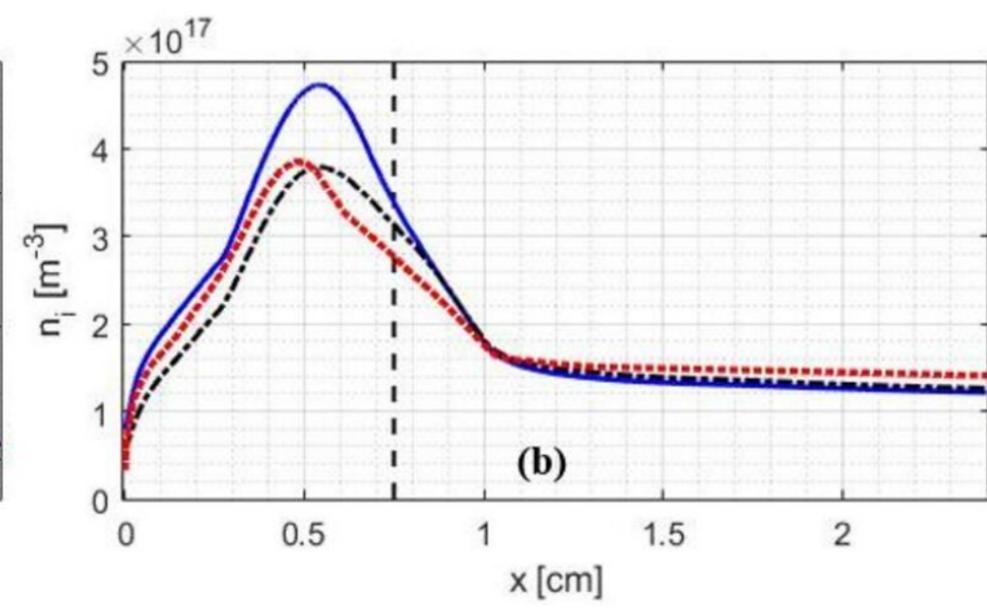

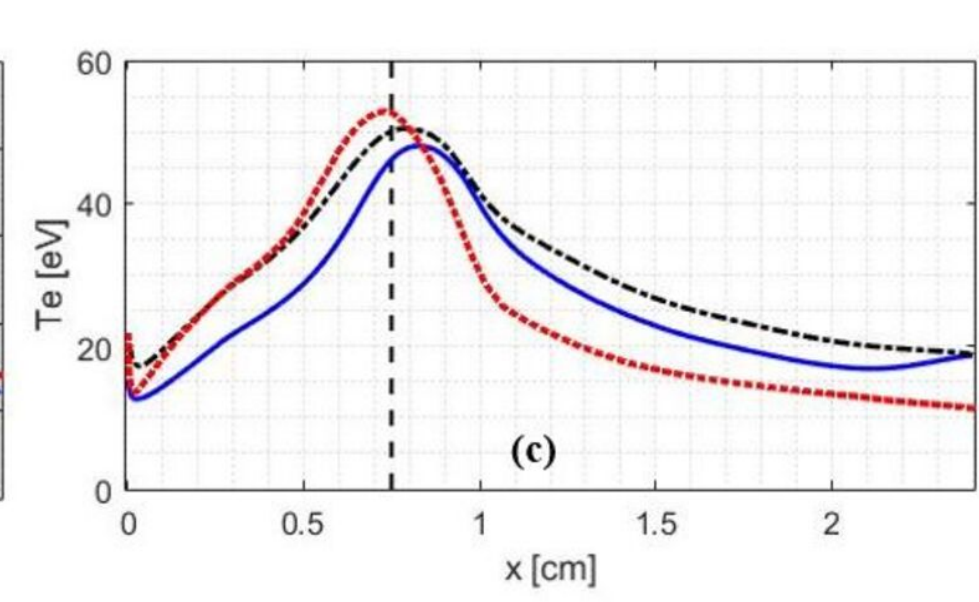

Single − region

Double − region

2D simulations

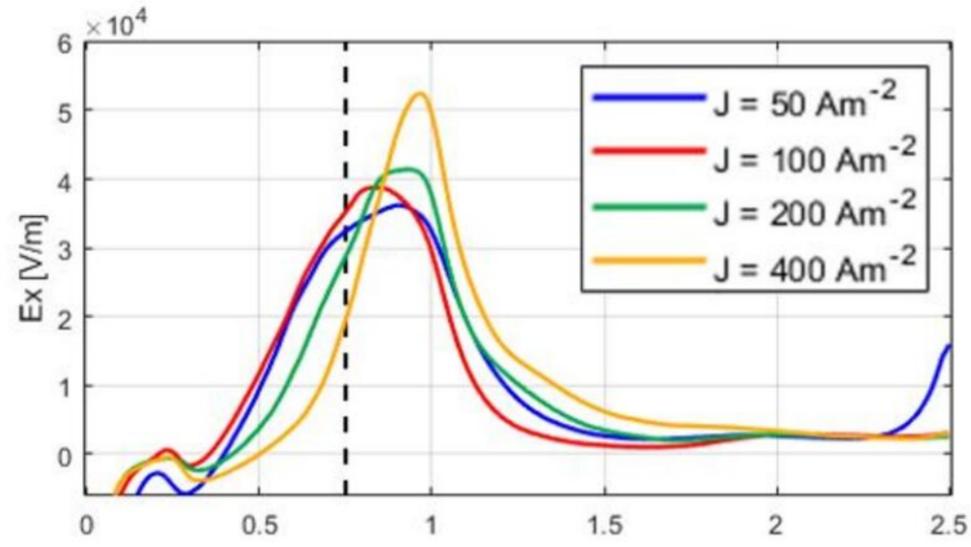

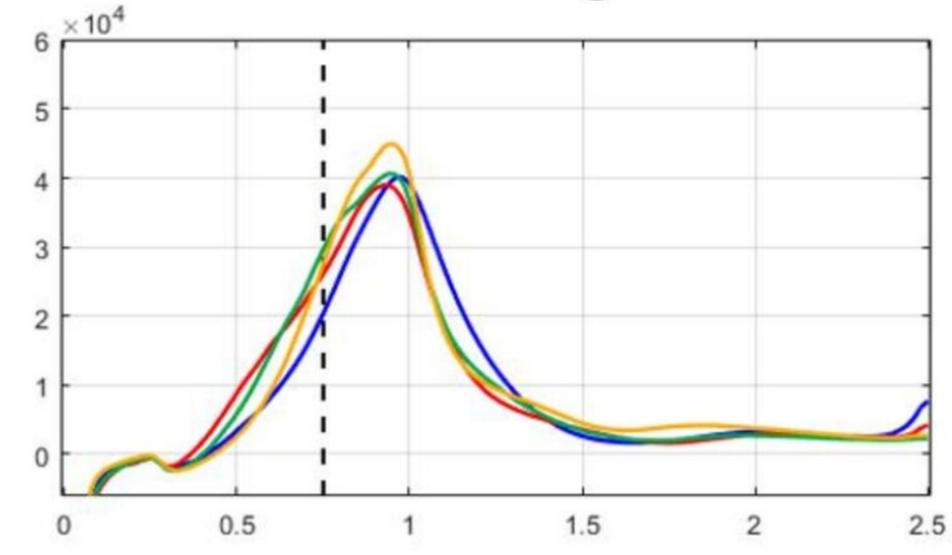

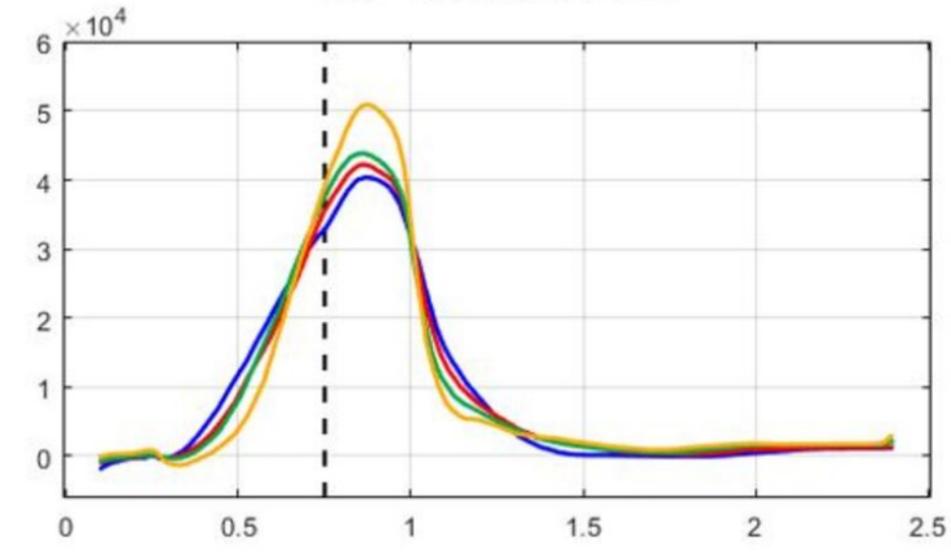

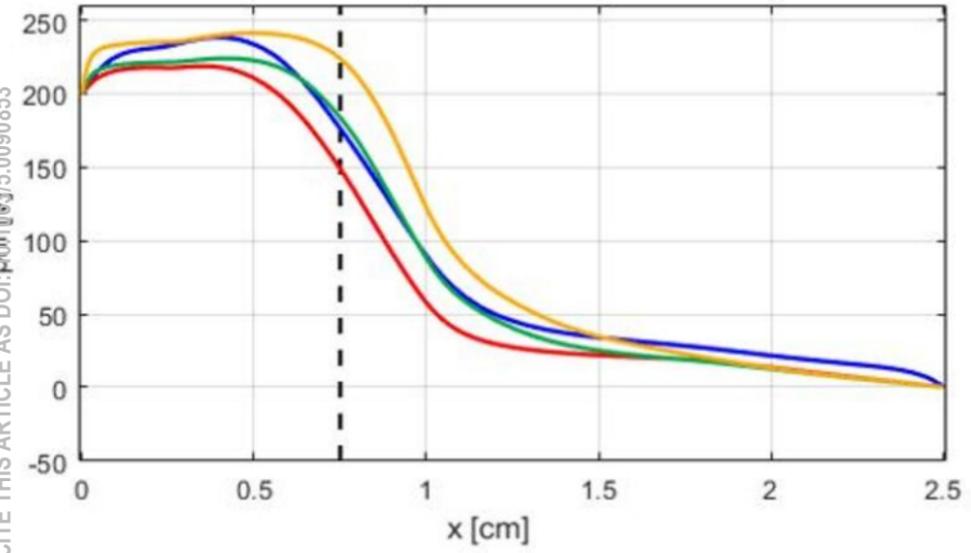

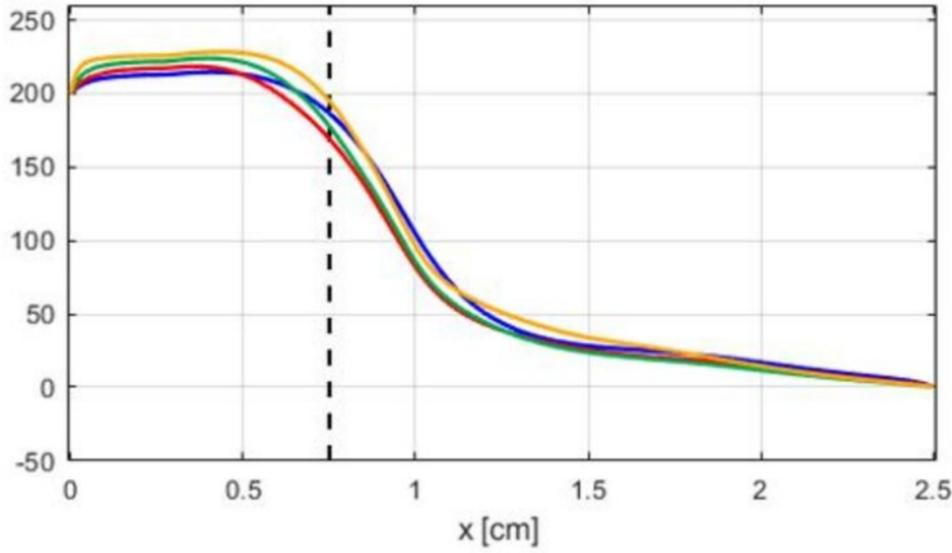

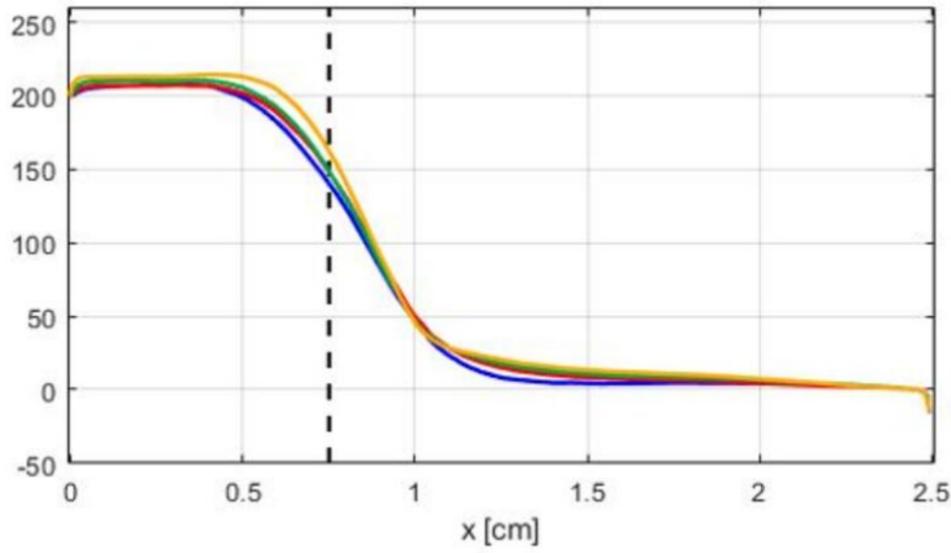

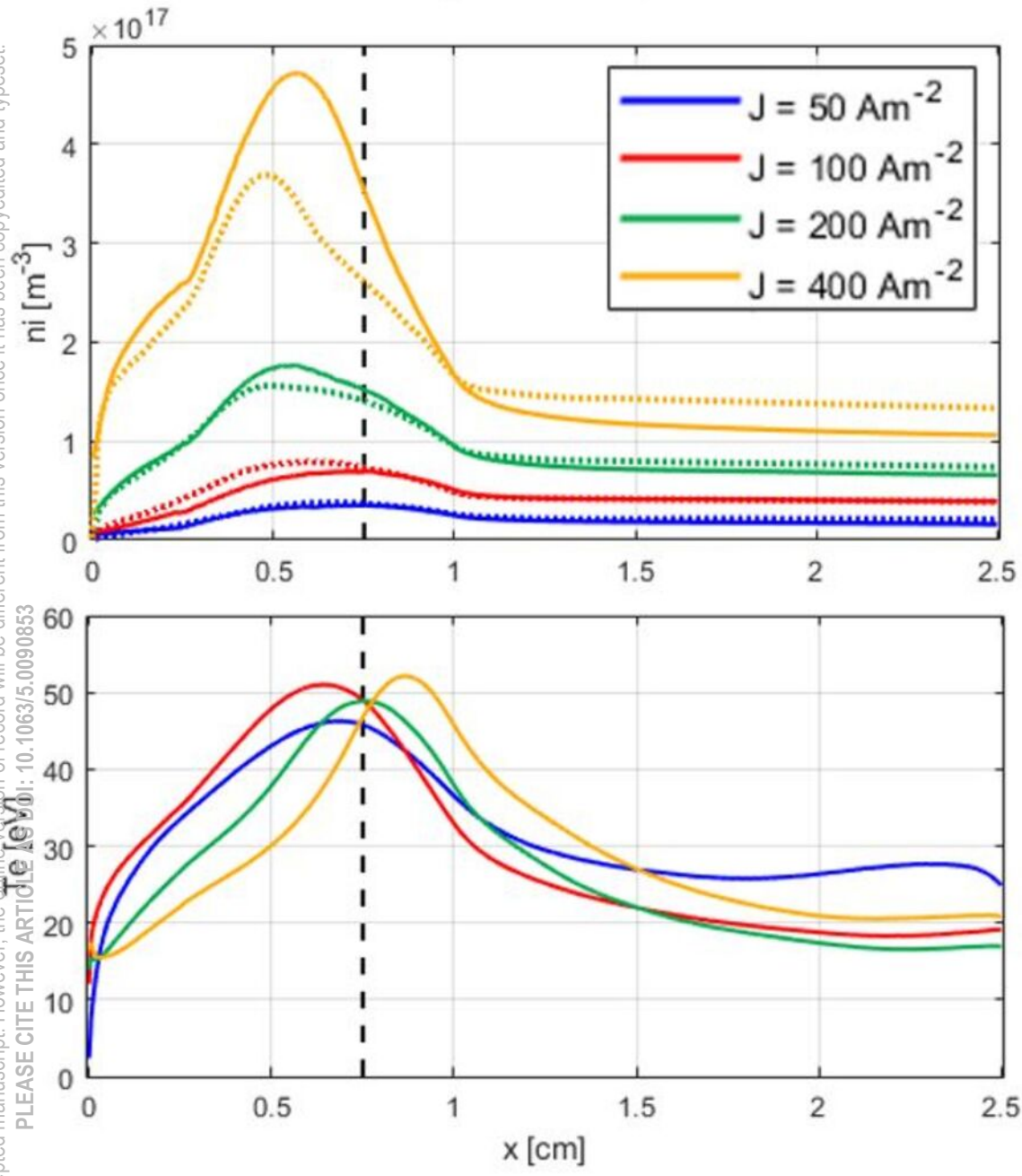
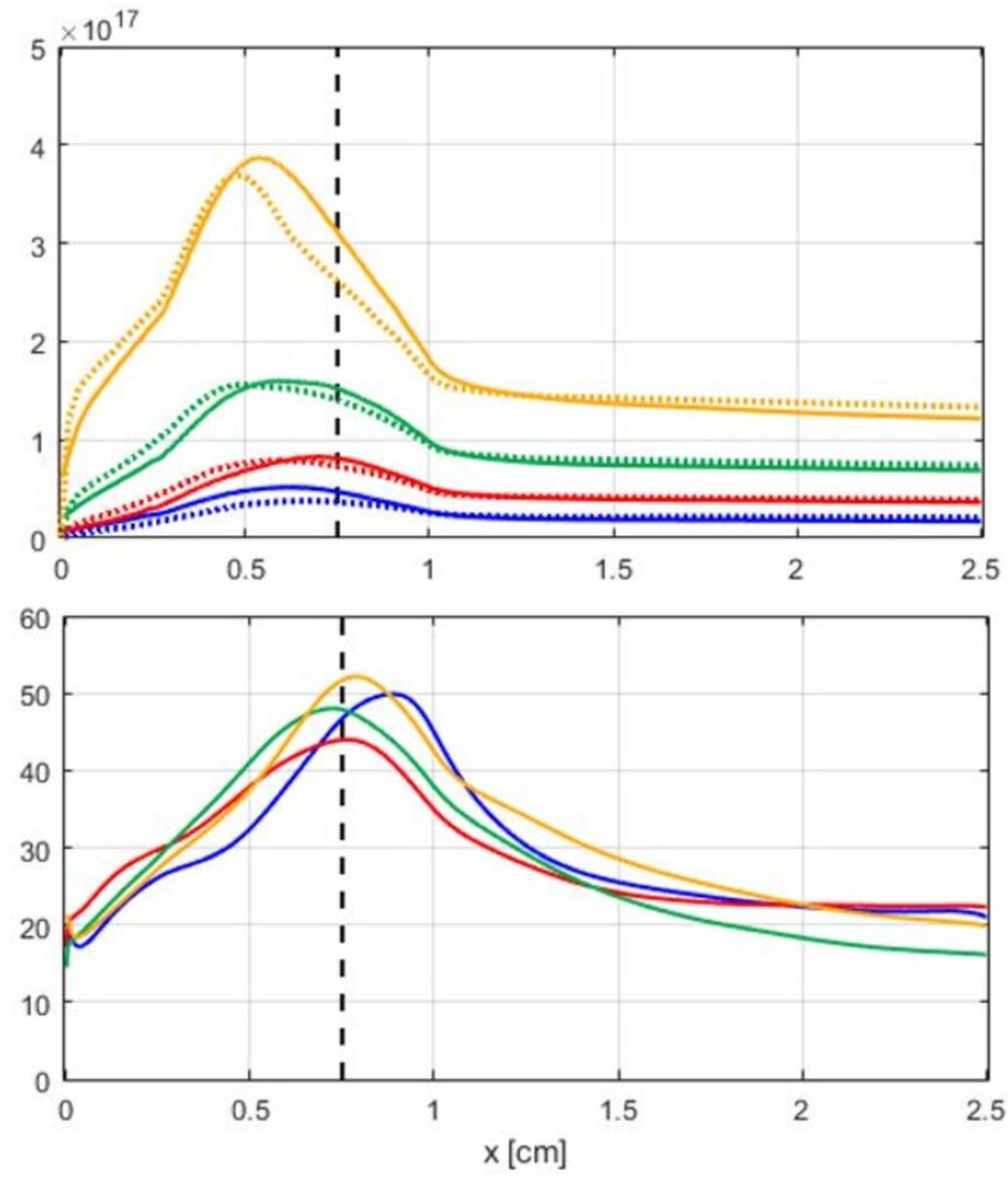

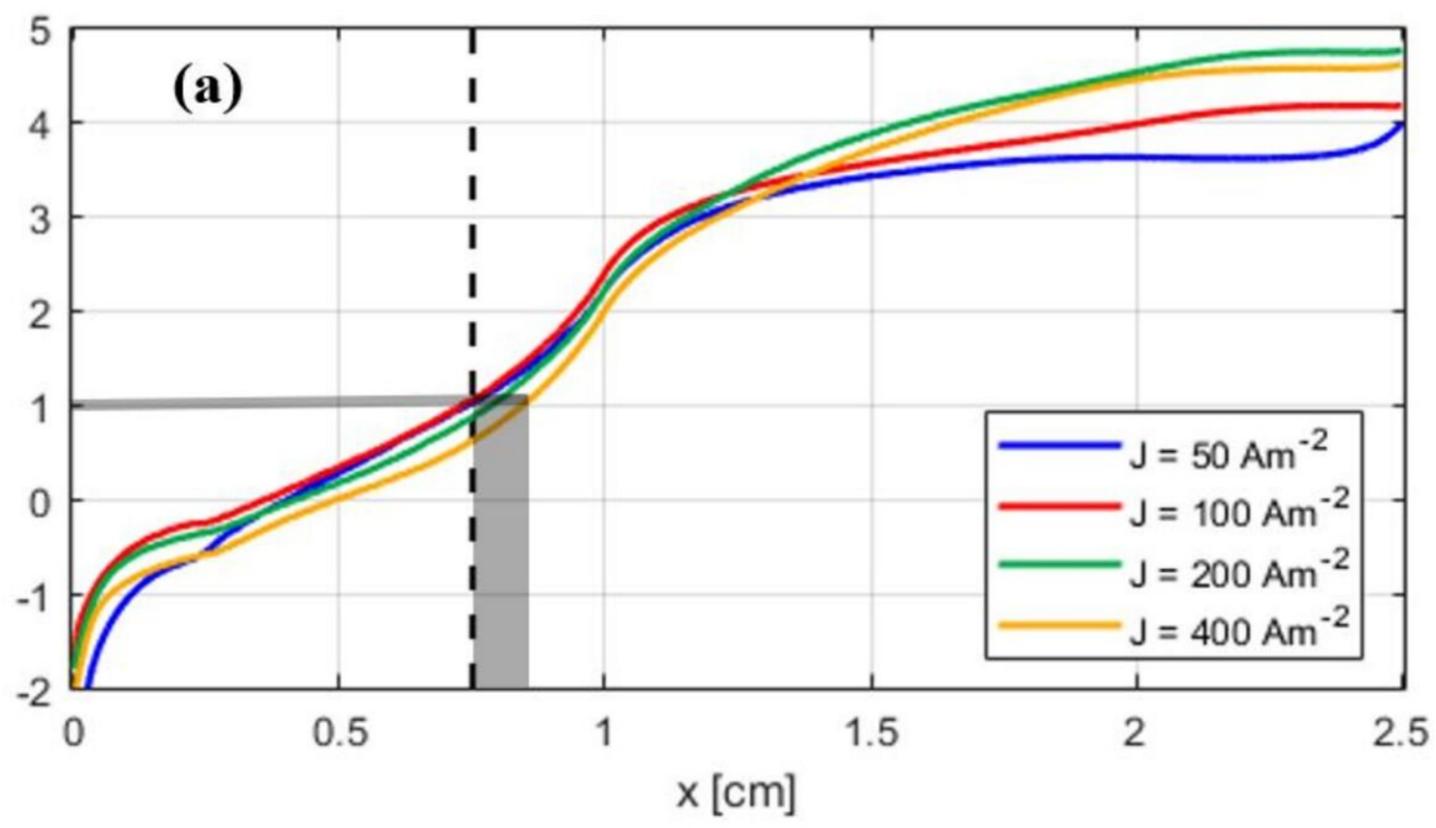

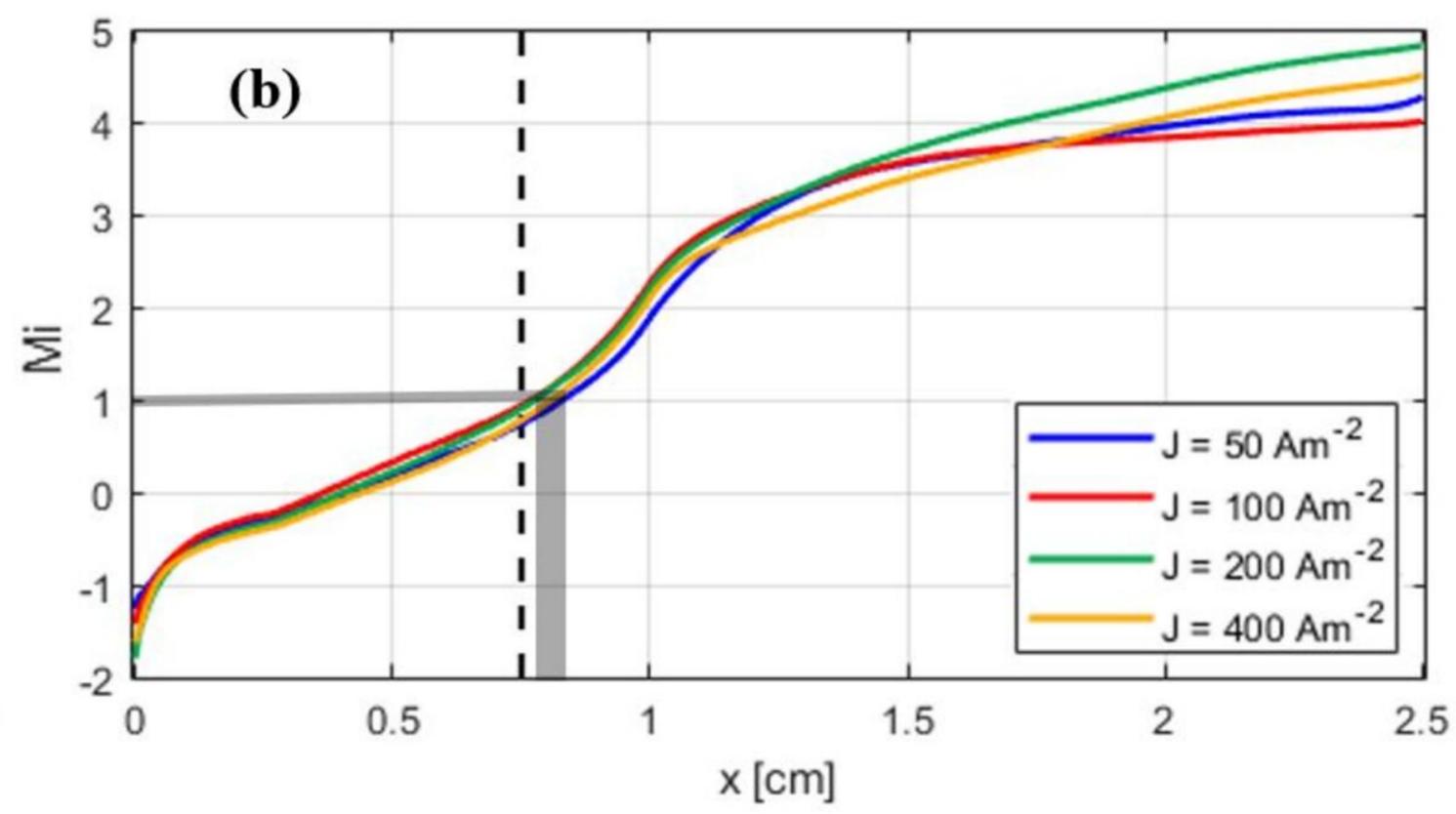

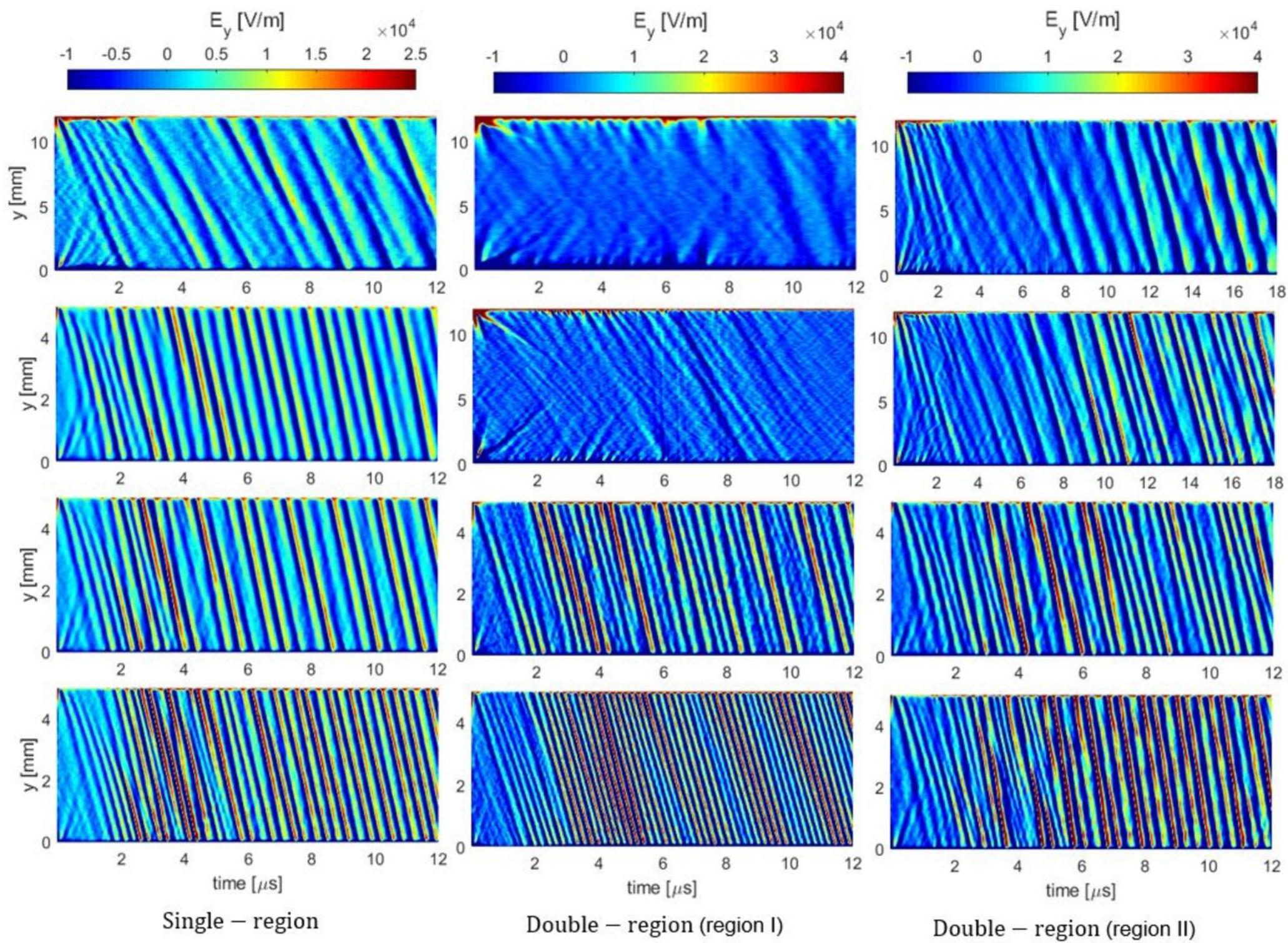

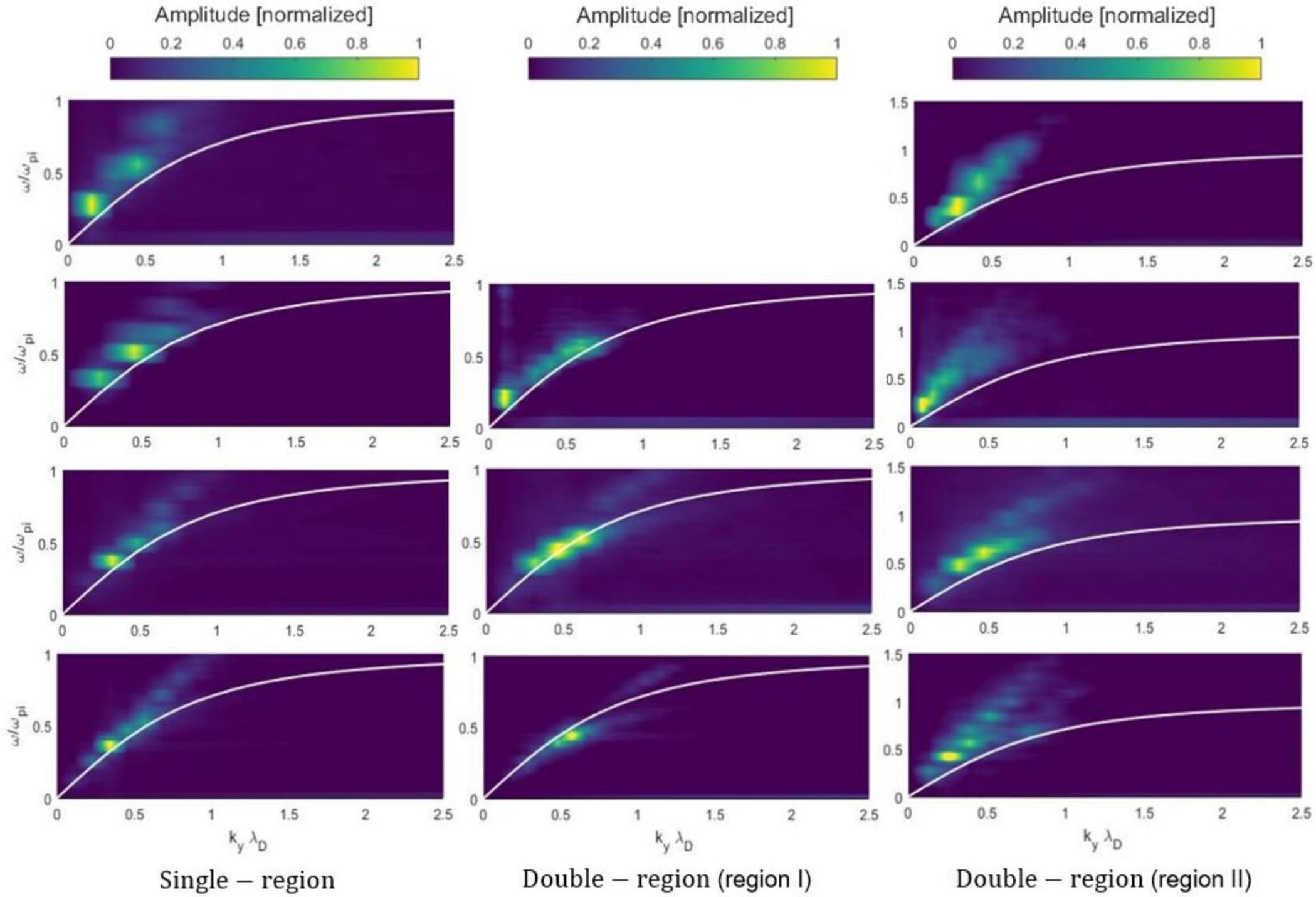

Single − region                    Double − region (region I)                    Double − region (region II)

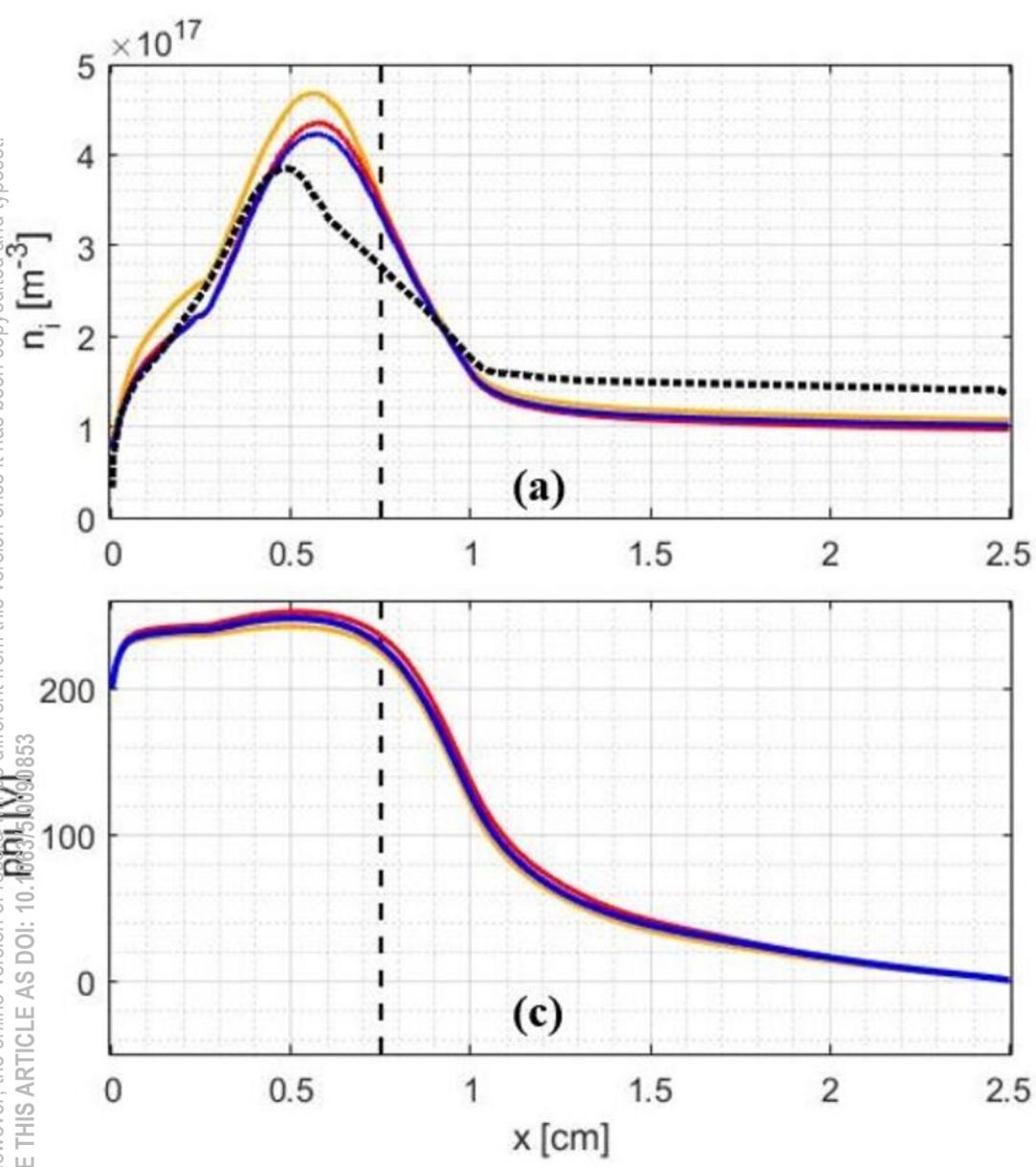

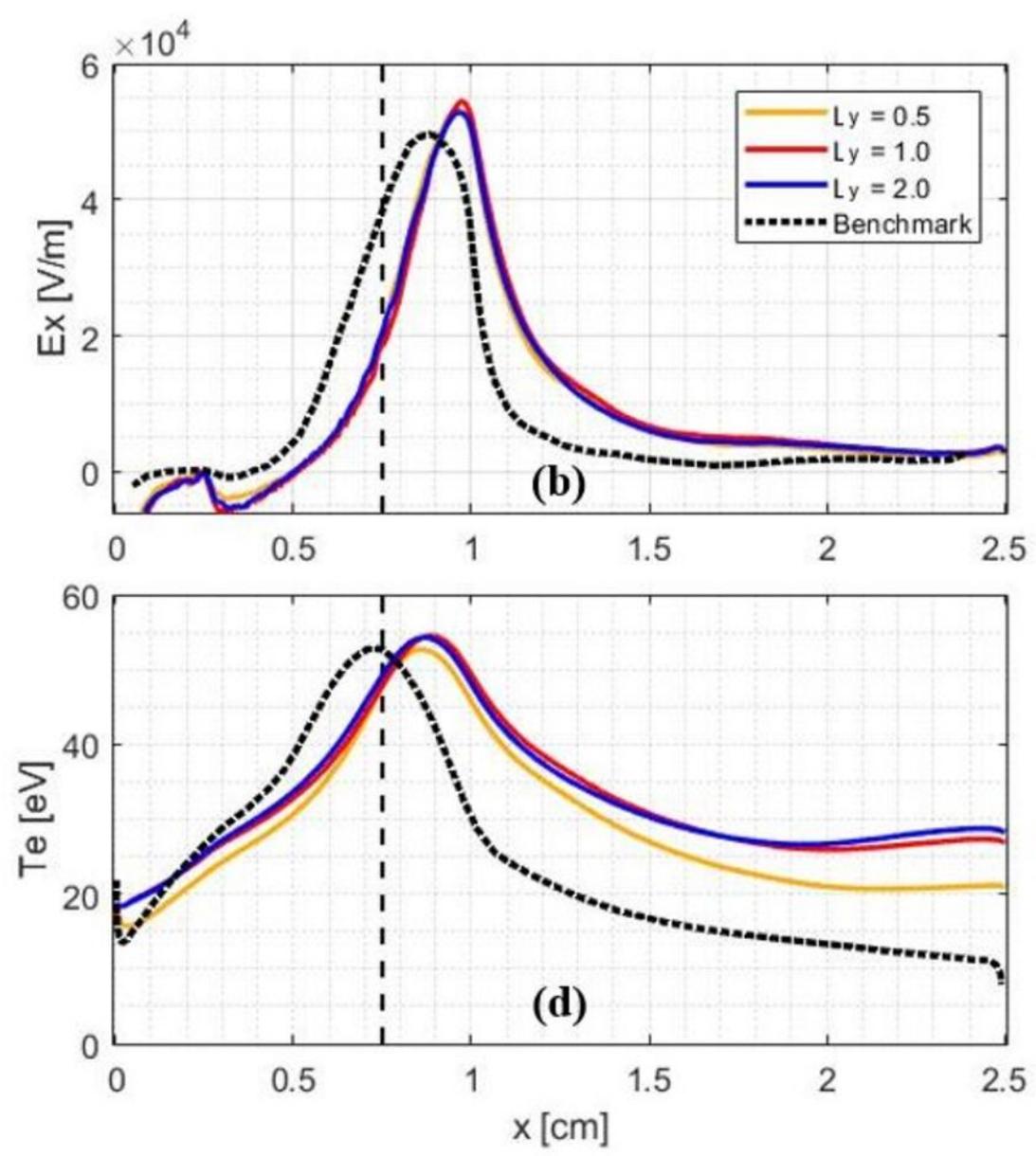

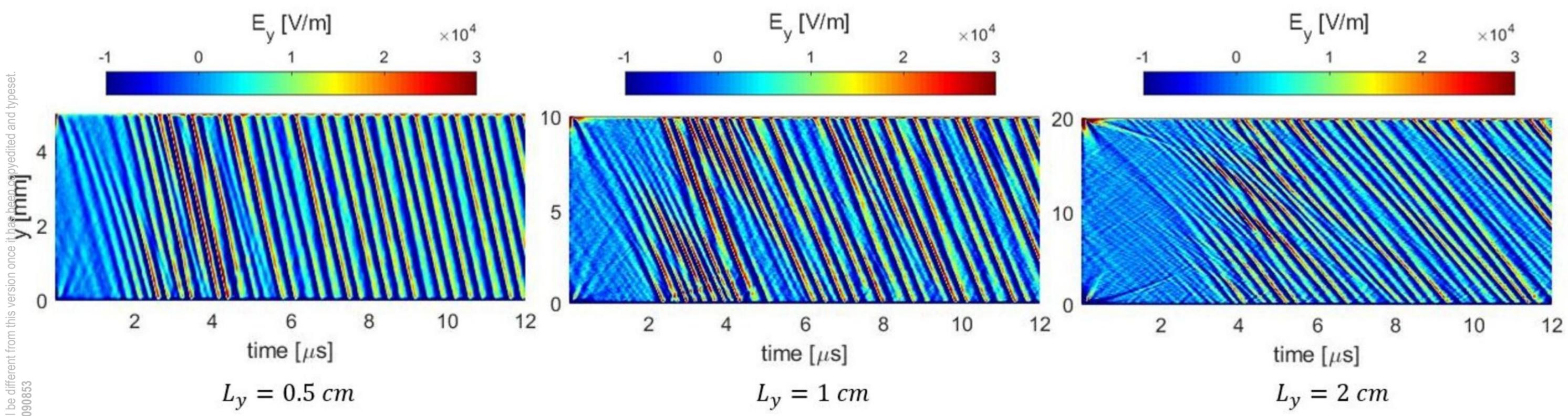

$L_y = 0.5 \ cm$ $\qquad\qquad\qquad L_y = 1 \ cm$ $\qquad\qquad\qquad L_y = 2 \ cm$

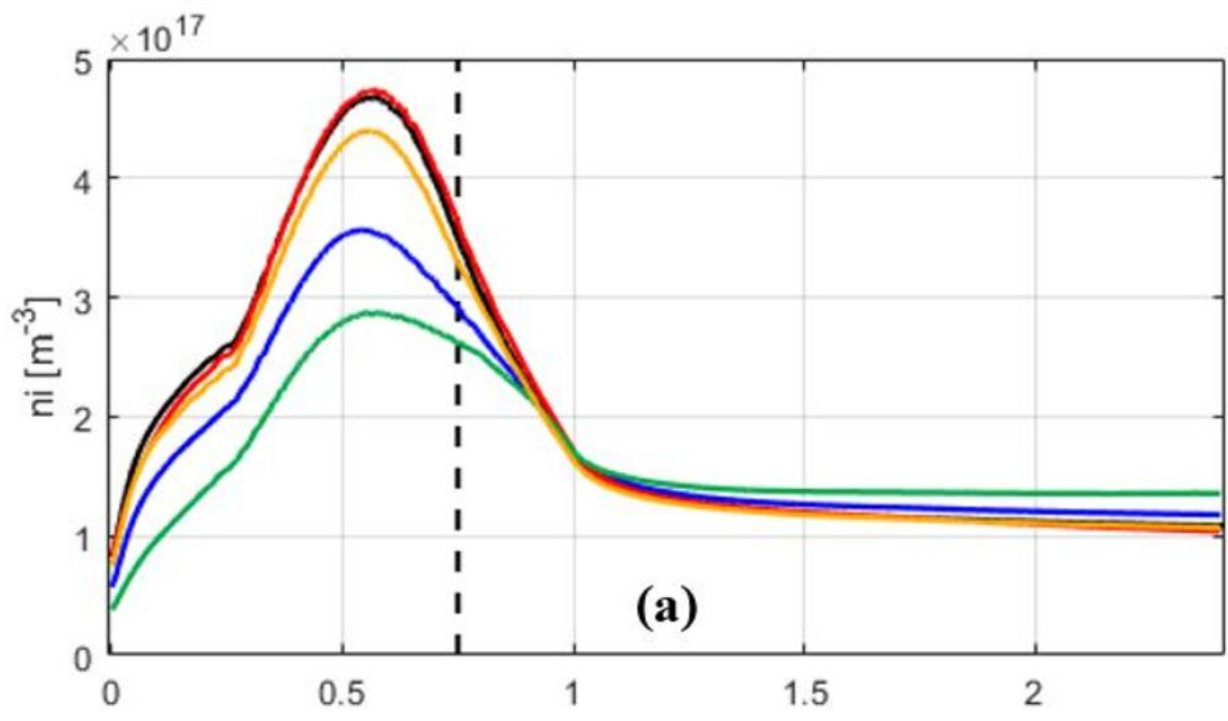
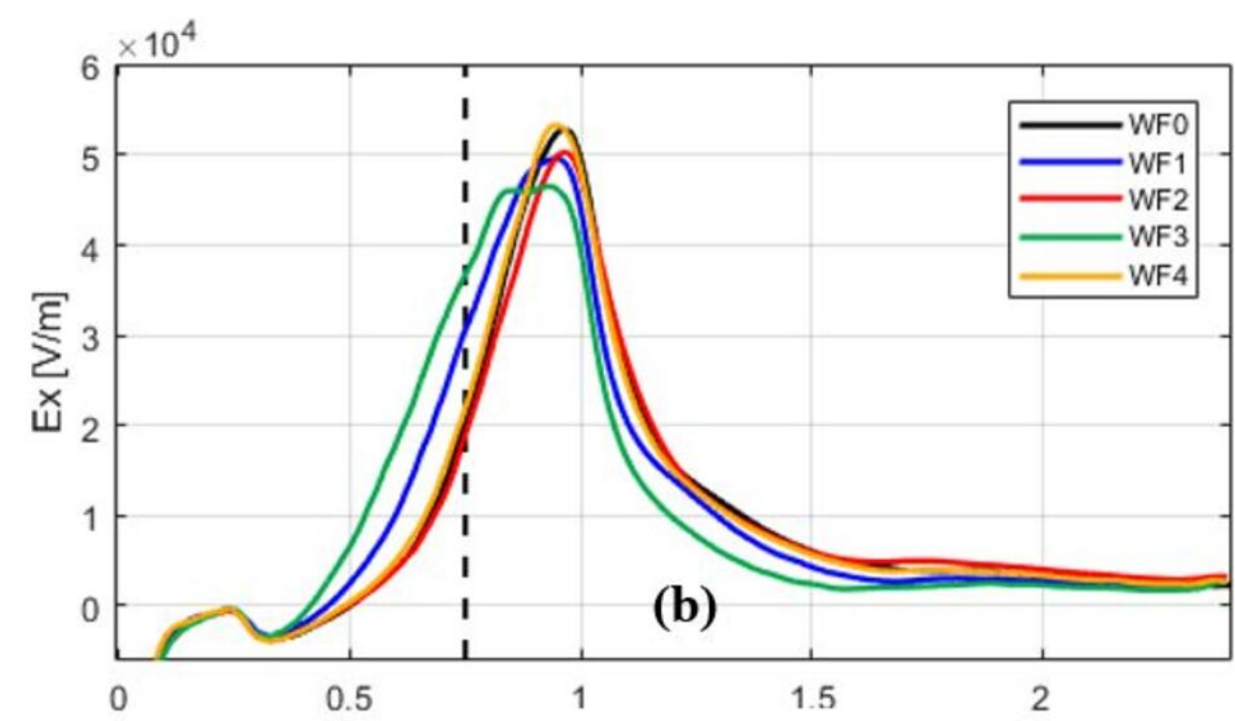
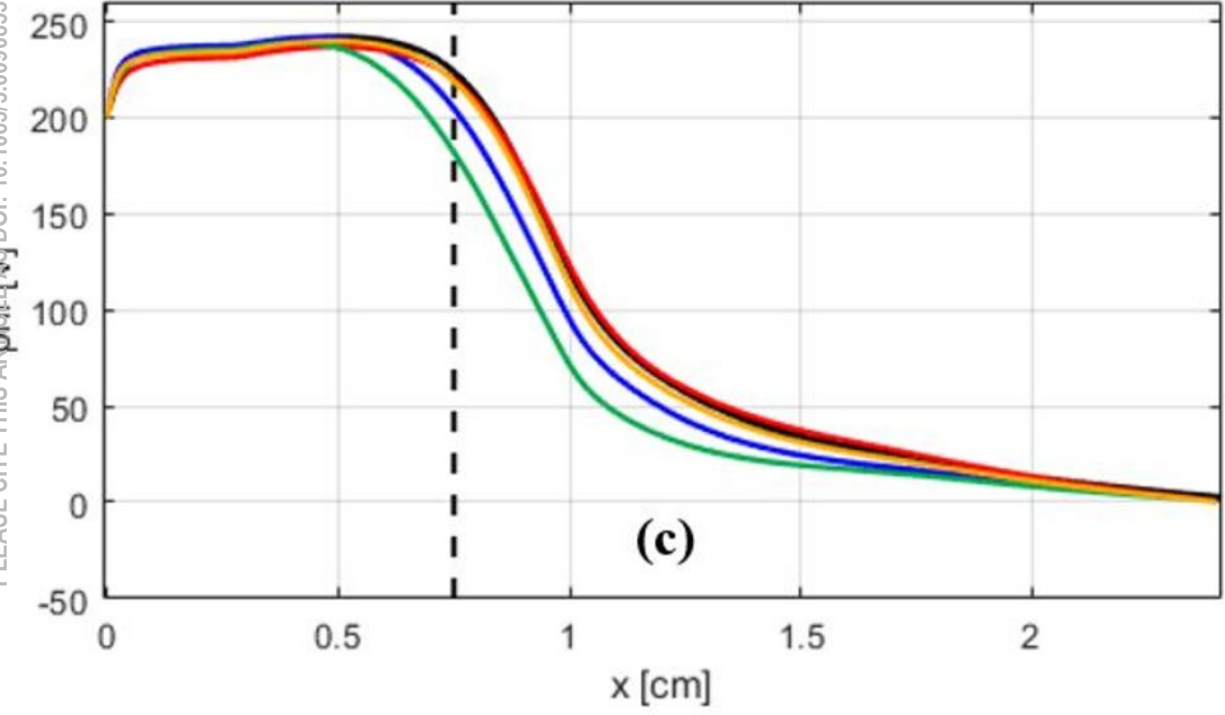
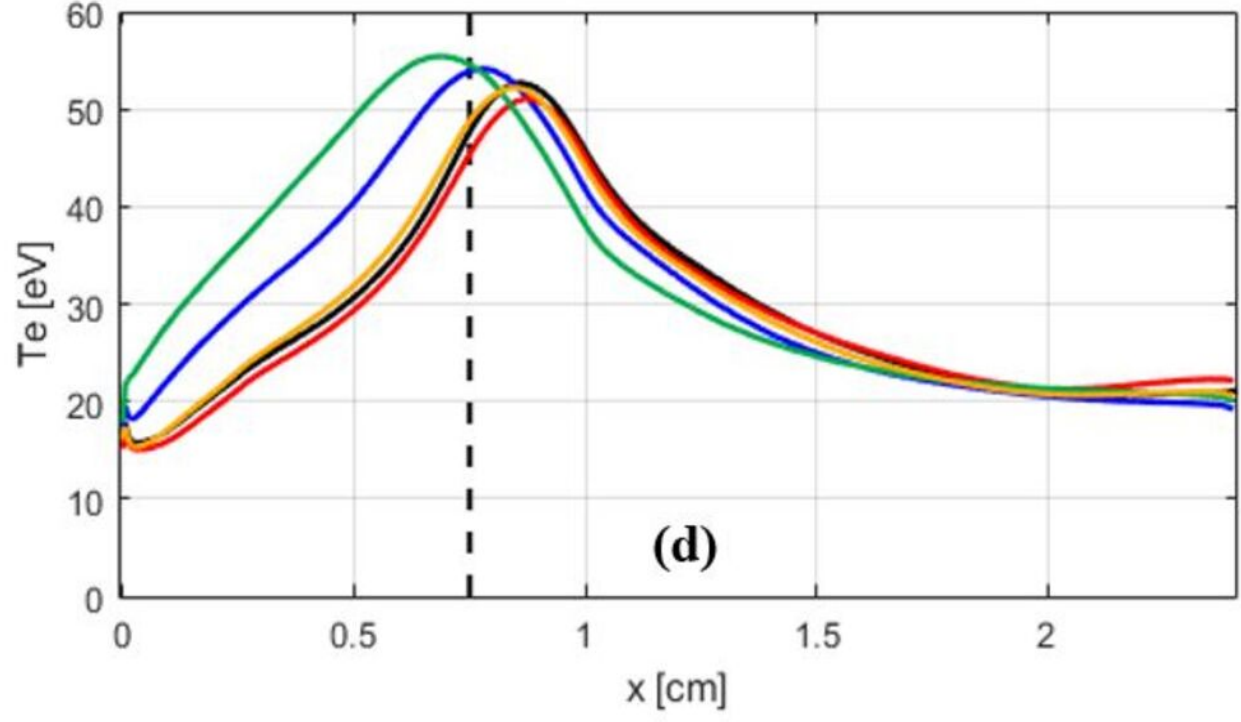

E$_y$ [V/m]  ×10⁴

Amplitude [normalized]

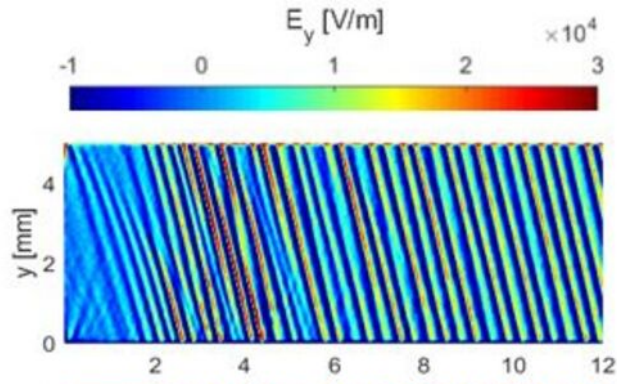 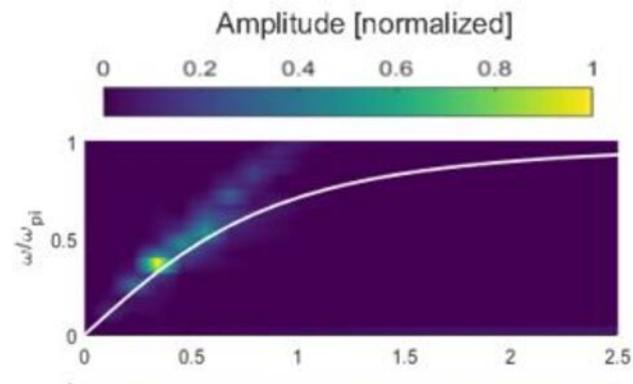

*WF*0

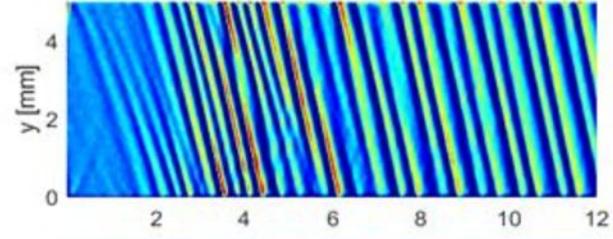 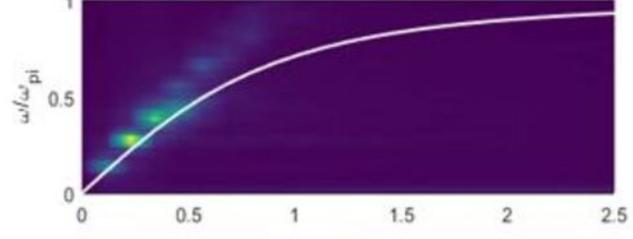

*WF*1

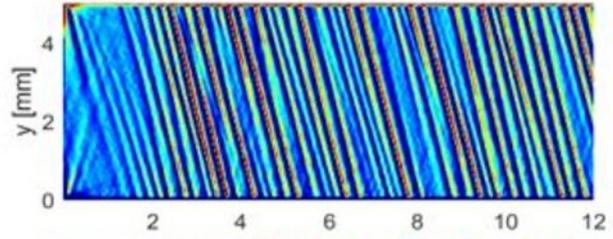 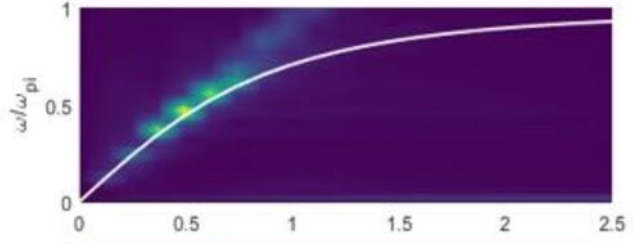

*WF*2

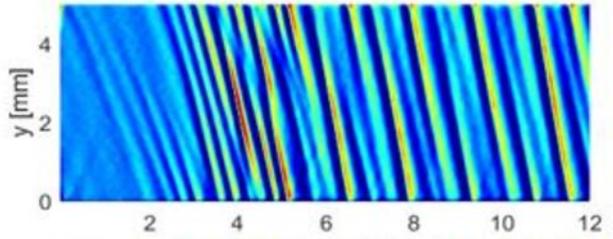 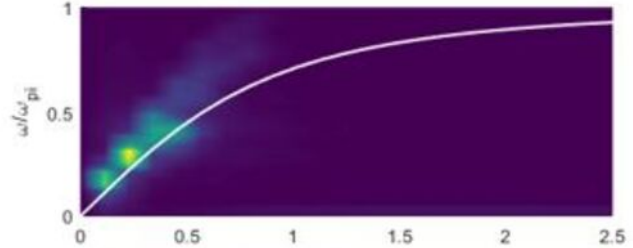

*WF*3

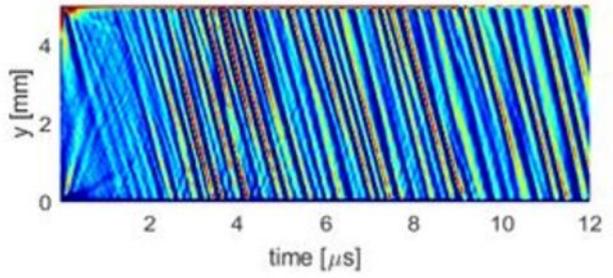 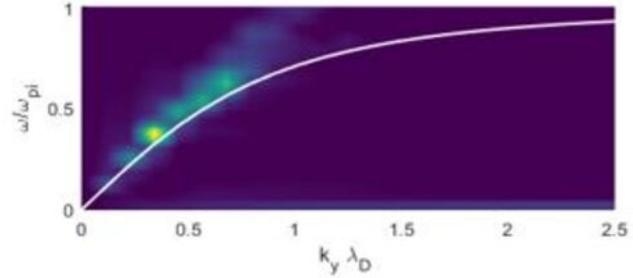

*WF*4

time [μs]

$k_y \lambda_D$

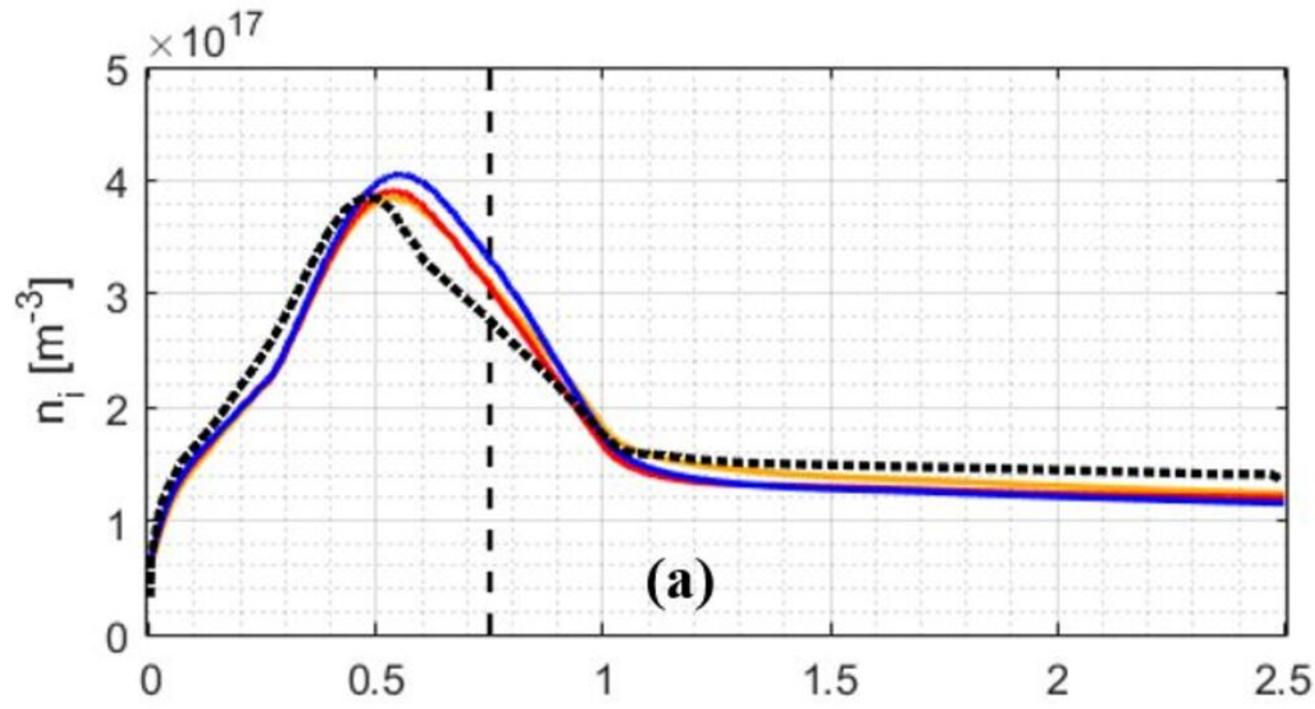

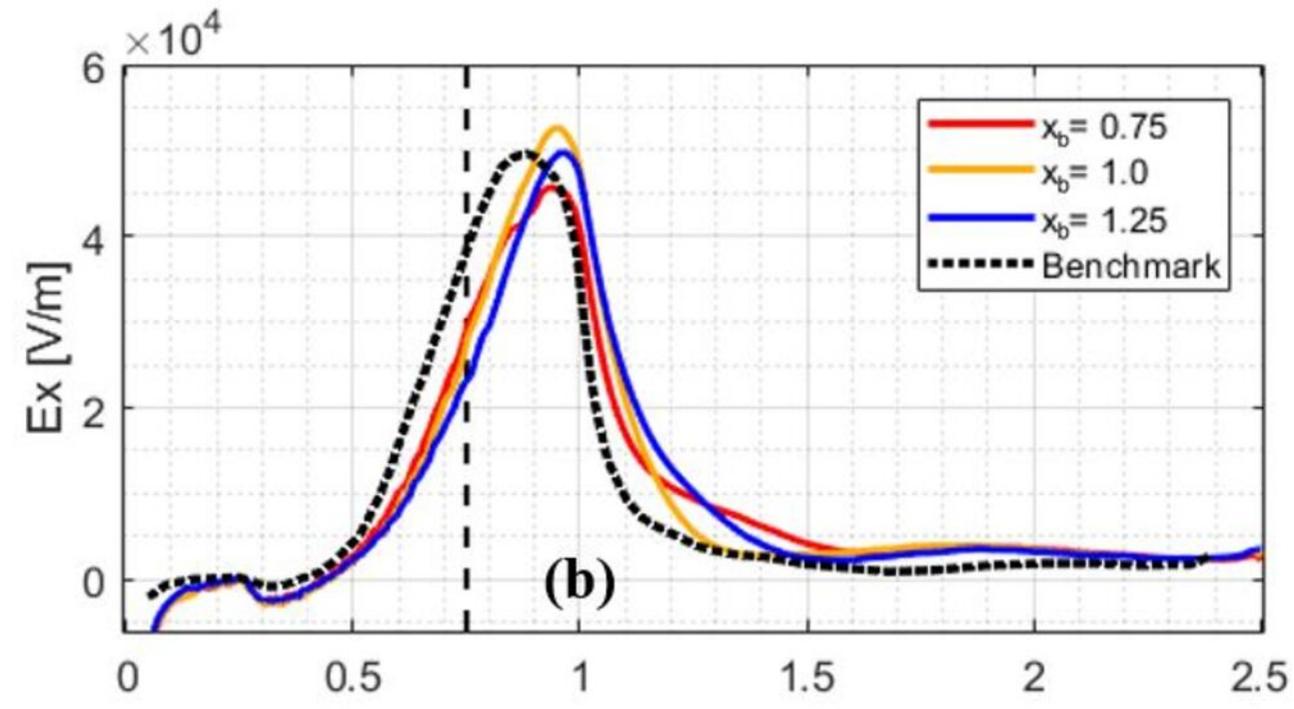

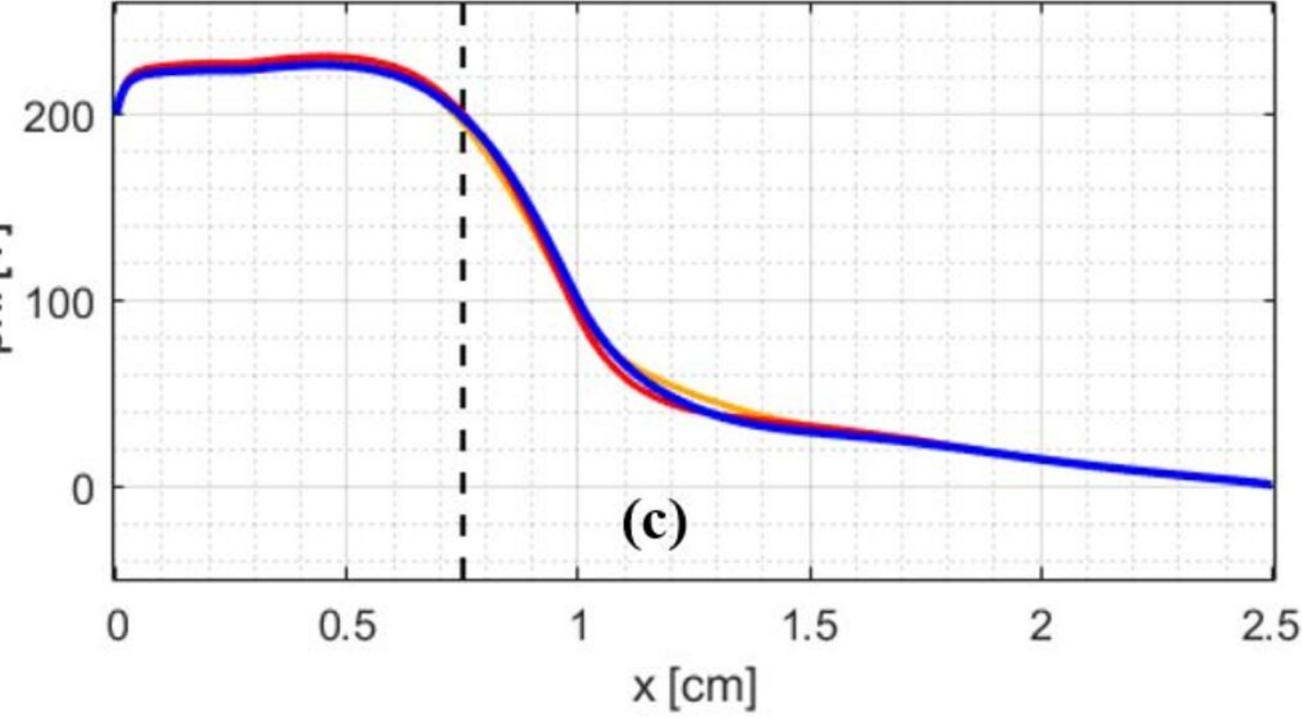

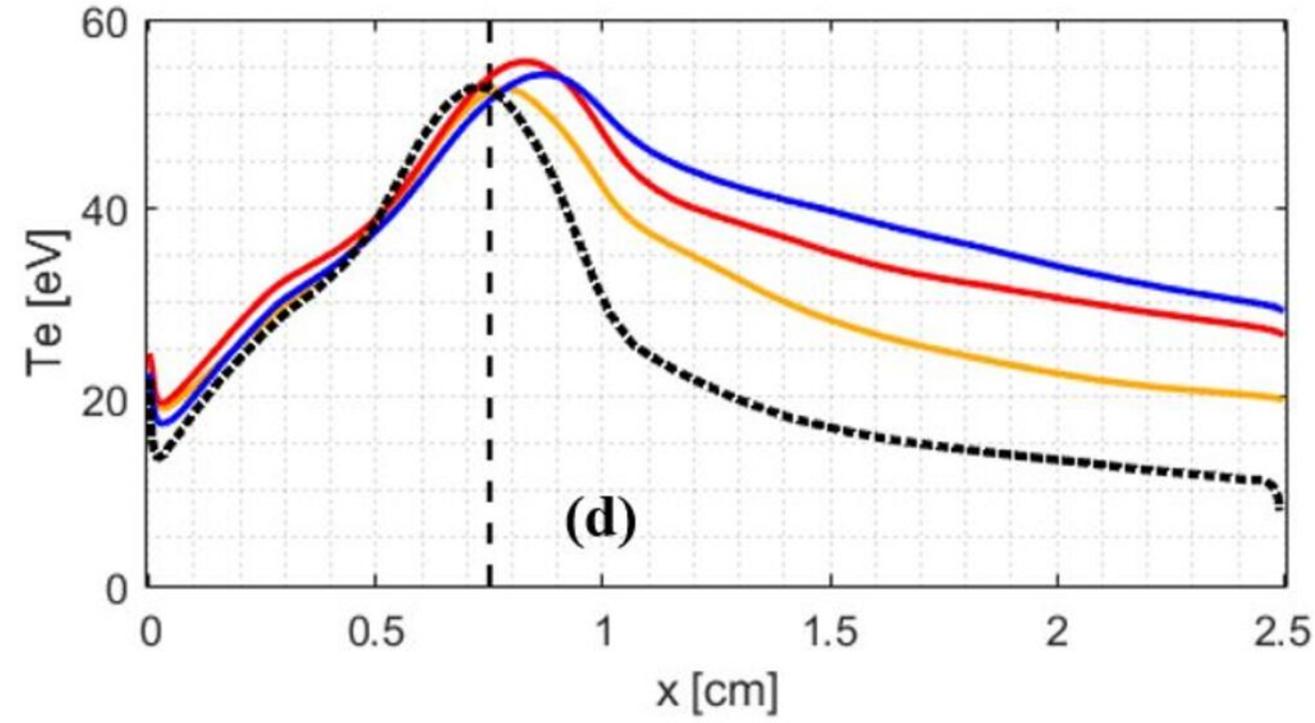

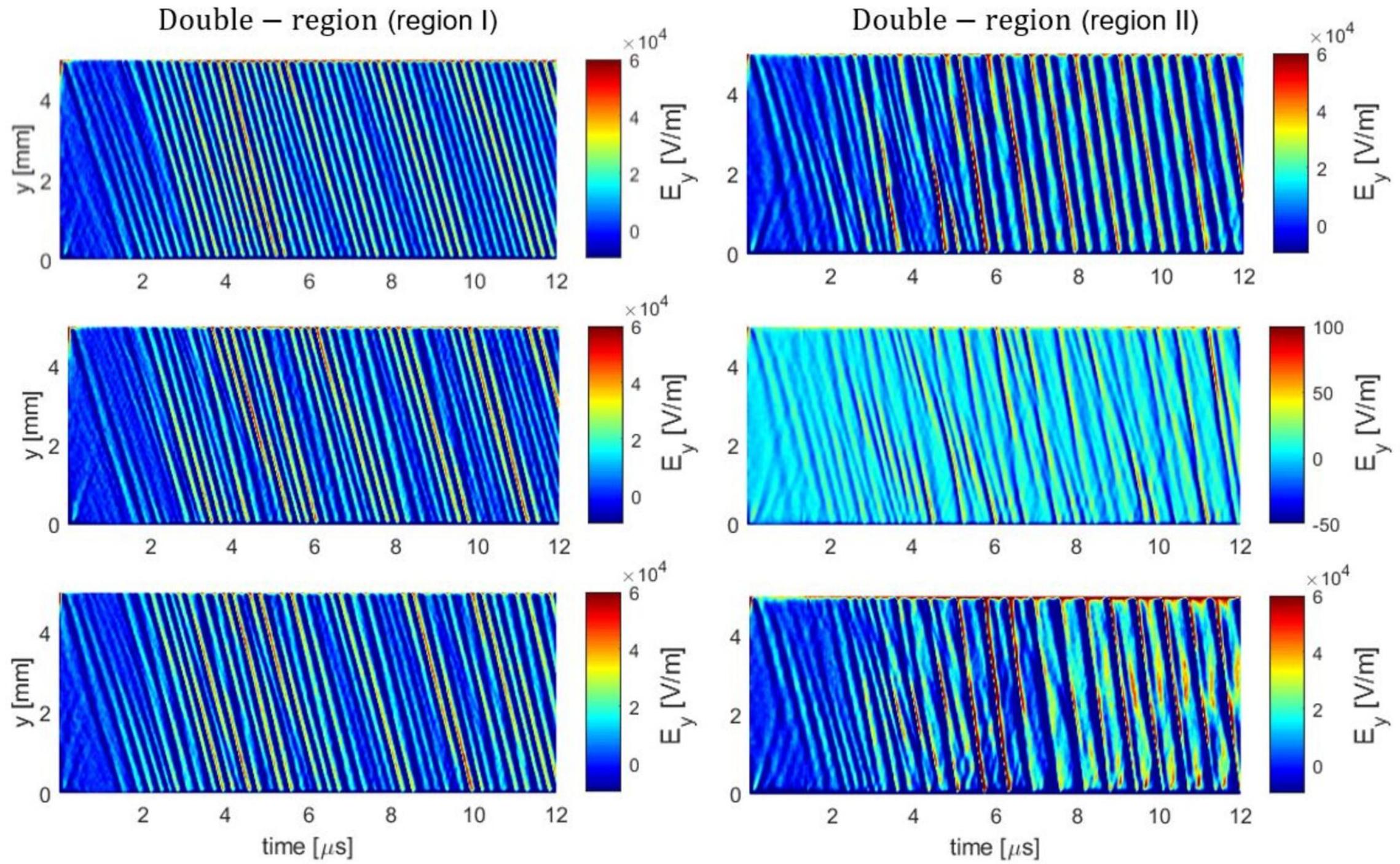

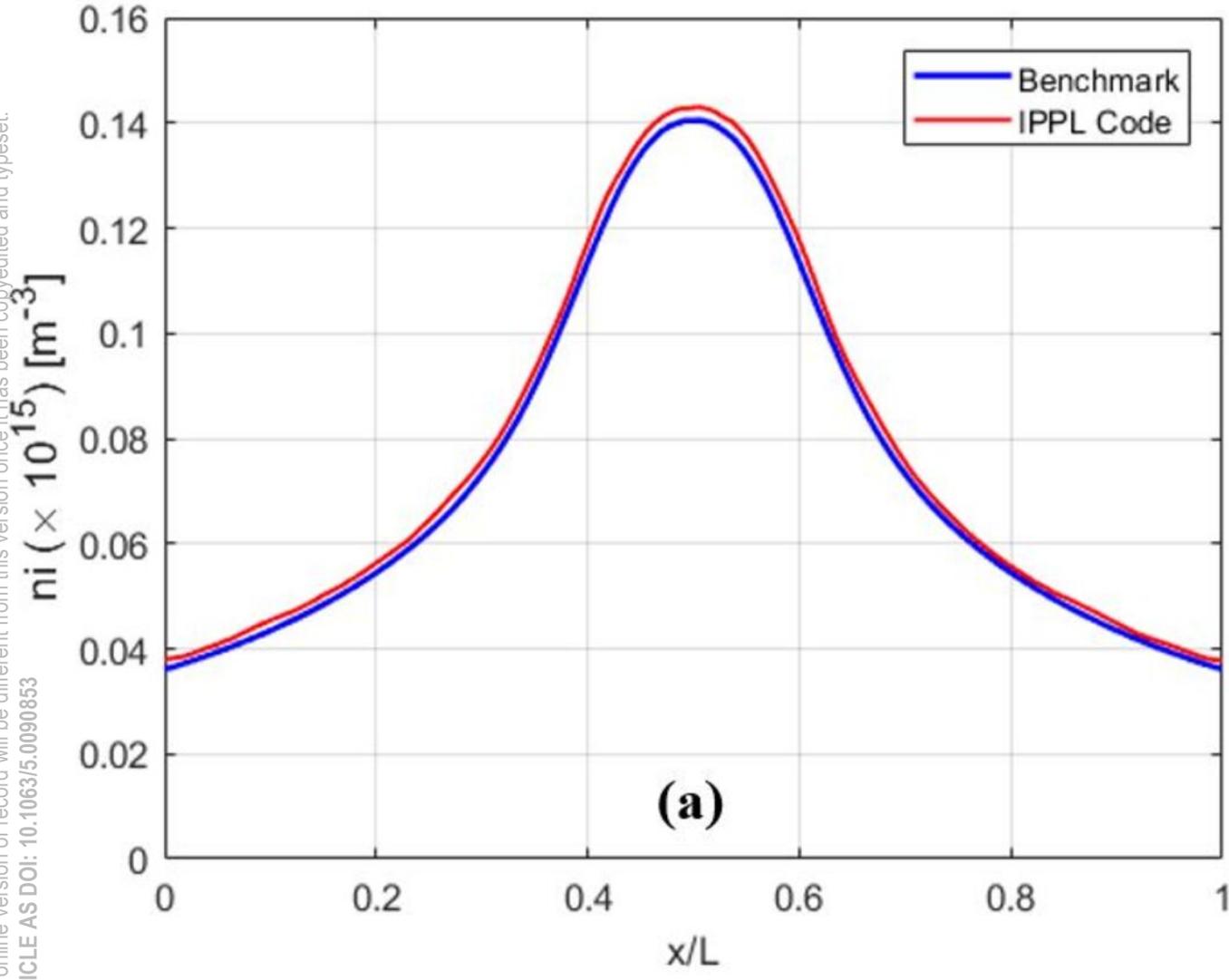

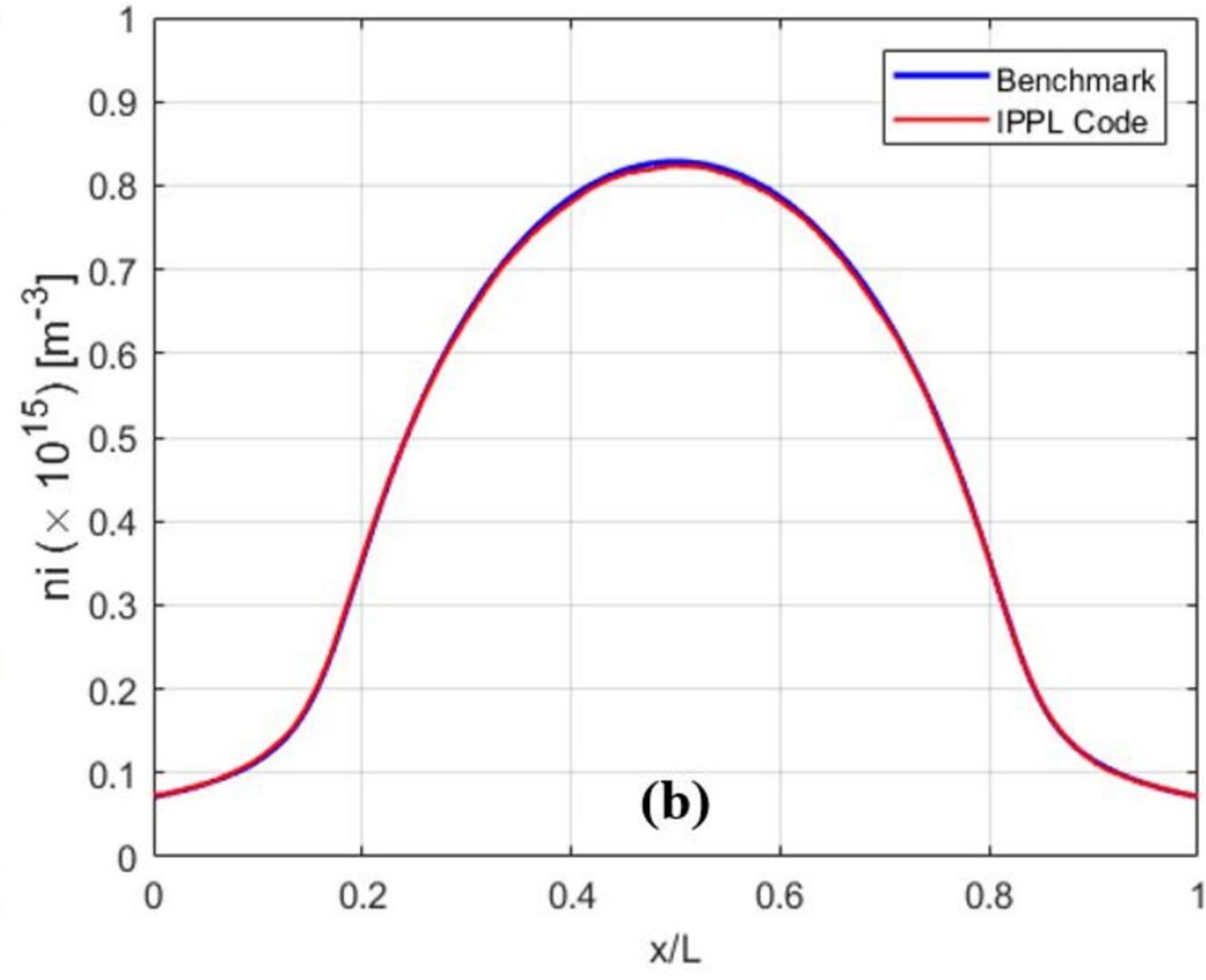

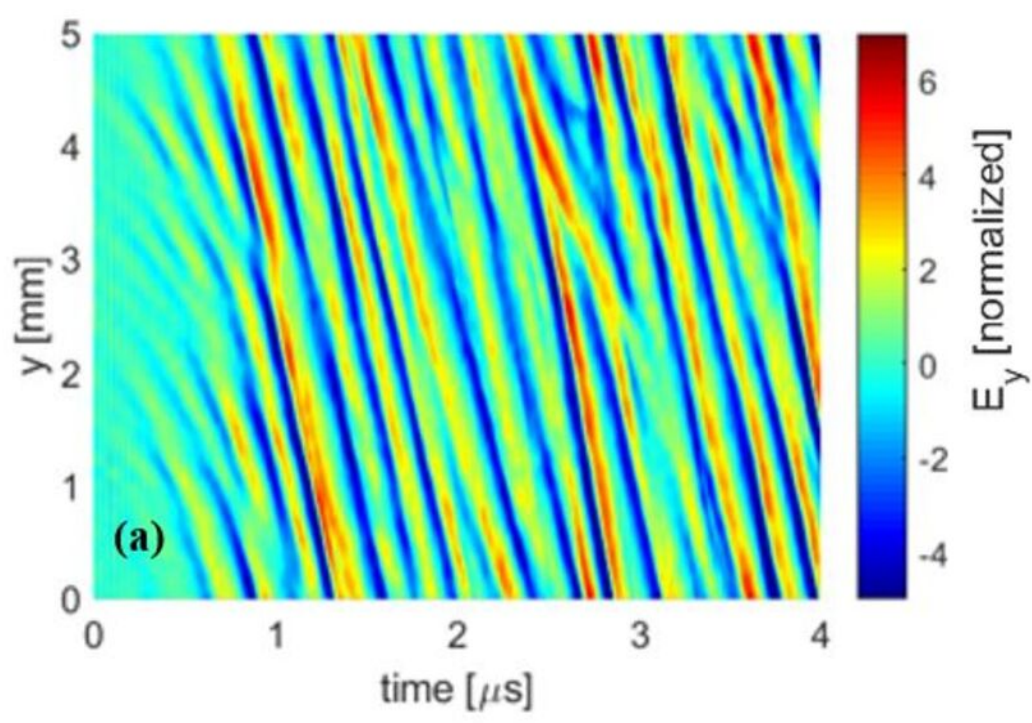

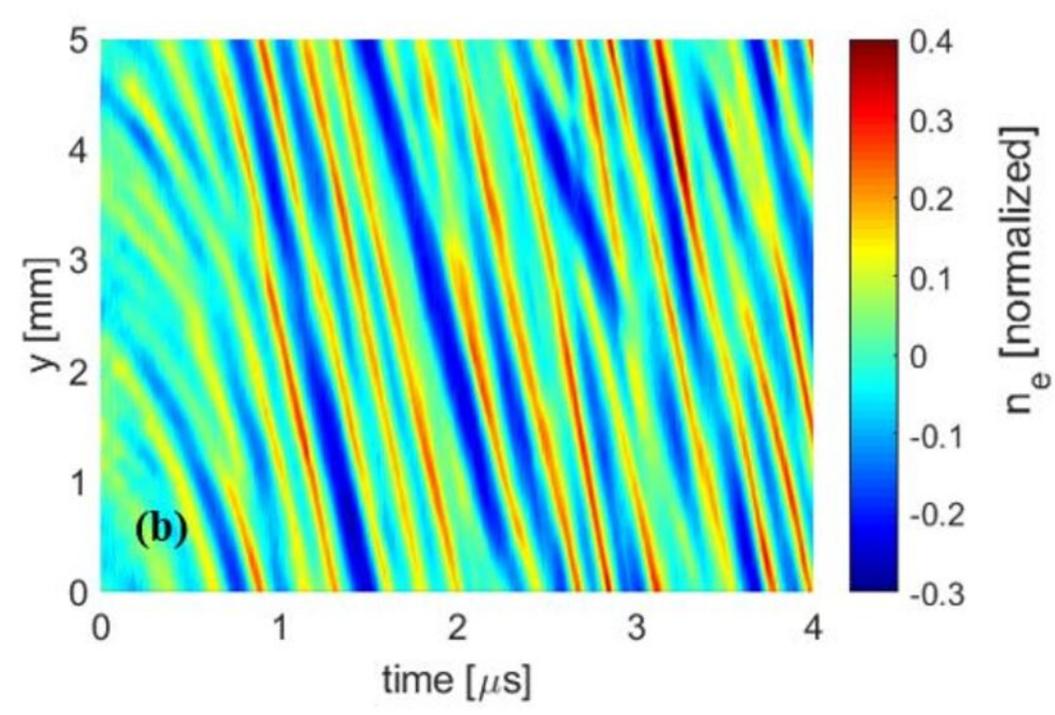

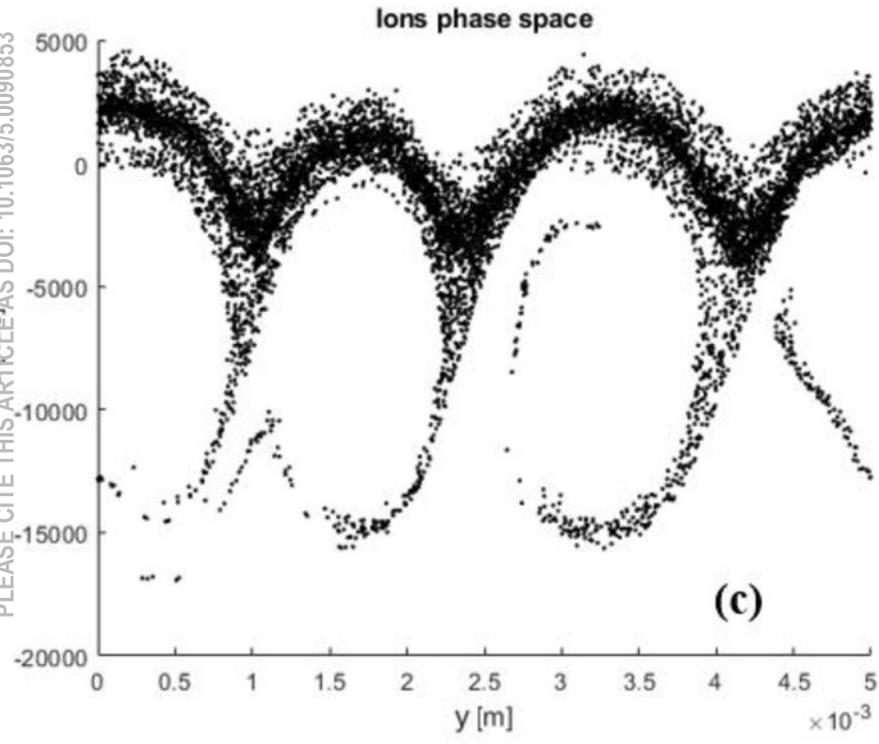

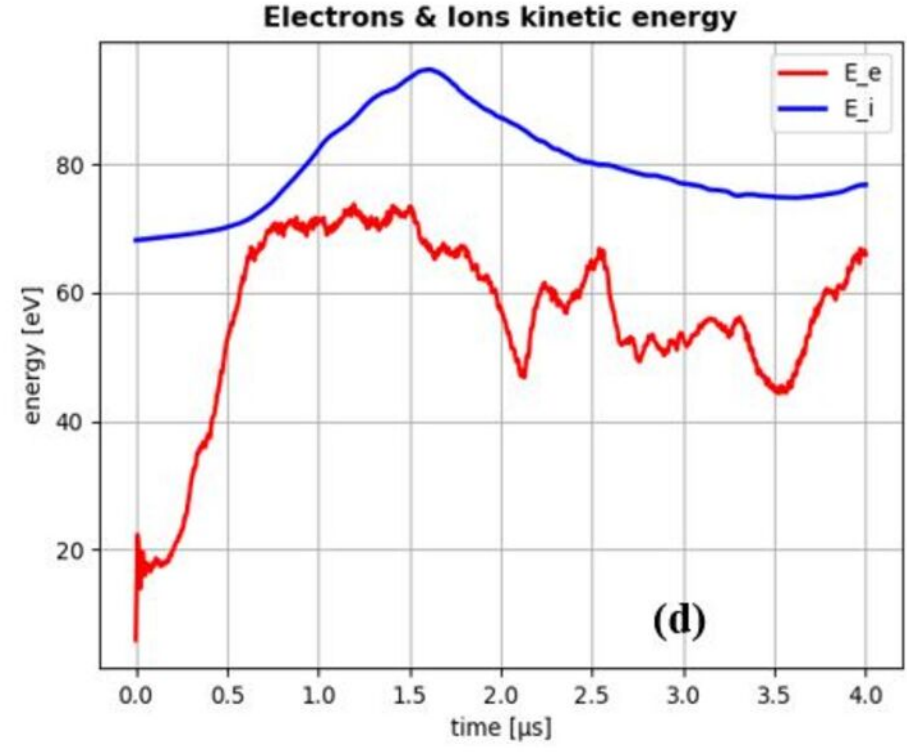

Double − region (region I) and Double − region (region II) plots of $E_y$ [V/m] versus time [$\mu$s] and $y$ [mm] for $x_b = 0.75\ cm$, $x_b = 1\ cm$, and $x_b = 1.25\ cm$.